\documentclass{IEEEtran}
\usepackage{cite}
\usepackage{amsmath,amssymb,amsfonts}
\usepackage{algorithmic}
\usepackage{graphicx,color}
\usepackage{textcomp}
\usepackage{xcolor}
\usepackage{hyperref}
\usepackage{amsmath,amsfonts}
\usepackage{array}
\usepackage[caption=false,font=normalsize,labelfont=sf,textfont=sf]{subfig}
\usepackage{textcomp}
\usepackage{stfloats}
\usepackage{url}
\usepackage[caption=false]{subfig}
\usepackage[justification=centering]{caption}
\usepackage{verbatim}
\usepackage{graphicx}
\usepackage{xcolor}
\usepackage{amssymb}
\usepackage{cite}
\usepackage[ruled,vlined]{algorithm2e}
\def\BibTeX{{\rm B\kern-.05em{\sc i\kern-.025em b}\kern-.08em
    T\kern-.1667em\lower.7ex\hbox{E}\kern-.125emX}}
\AtBeginDocument{\definecolor{ojcolor}{cmyk}{0.93,0.59,0.15,0.02}}

\begin{document}


\title{6G goal-oriented communications:\\ How to coexist with legacy systems?}

\author{Mattia Merluzzi,~\IEEEmembership{Member,~IEEE}, Miltiadis C. Filippou,~\IEEEmembership{Senior Member,~IEEE},\\ 
Leonardo Gomes Baltar,~\IEEEmembership{Senior Member,~IEEE}, Markus D. Mück,~\IEEEmembership{Member,~IEEE}, \\ and Emilio Calvanese Strinati,~\IEEEmembership{Member,~IEEE}\IEEEcompsocitemizethanks{\IEEEcompsocthanksitem M.~Merluzzi and E. Calvanese Strinati are with Univ. Grenoble Alpes, CEA, Leti, F-38000 Grenoble, France, France. \\Email: mattia.merluzzi@cea.fr, emilio.calvanese-strinati@cea.fr.\IEEEcompsocthanksitem Miltiadis C. Filippou was with Intel Deutschland GmbH, 85579 Neubiberg, Germany, when the paper was prepared, he is now with Nokia Bell Labs, Munich, Germany. Email: miltiades.filippou.de@ieee.org.  \IEEEcompsocthanksitem Leonardo Gomes Baltar and Markus D. Mück are with Intel Deutschland GmbH, 85579 Neubiberg, Germany \\Email: leonardo.gomes.baltar@intel.com, markus.dominik.mueck@intel.com. \protect\\
This work was partly supported by the European Commission through the H2020 project Hexa-X (Grant Agreement no. 101015956)}}

\markboth{6G goal-oriented communications:  How to coexist with legacy systems?}{Mattia Merluzzi\textit{et al.}}
\maketitle
\begin{abstract}
6G will connect heterogeneous intelligent agents to make them natively operate complex cooperative tasks.
When connecting intelligence, two main research questions arise to identify how artificial intelligence and machine learning models behave depending on: \textit{i) }their input data quality, affected by errors induced by interference and additive noise during wireless communication; \textit{ii)} their contextual effectiveness and resilience to interpret and exploit the meaning behind the data. 
Both questions are within the realm of \textit{semantic} and \textit{goal-oriented} communications. 
With this paper we investigate how to effectively share communication spectrum resources between a legacy communication system (i.e., data-oriented) and a new goal-oriented edge intelligence one.
Specifically, we address the scenario of an enhanced Mobile Broadband (eMBB) service, i.e., a user uploading a video stream to a radio access point, interfering with an edge inference system, in which a user uploads images to a Mobile Edge Host that runs a classification task. Our objective is to achieve, through cooperation, the highest eMBB service data rate, subject to a targeted \textit{goal-effectiveness} of the edge inference service, namely the probability of confident inference on time. We first formalize a general definition of a goal in the context of wireless communications. This includes the goal-effectiveness, (i.e., the goal achievability rate, or the probability of achieving the goal), as well as that of \emph{goal cost} (i.e., the network resource consumption needed to achieve the goal with target effectiveness). We argue and show, through numerical evaluations, that communication reliability and goal-effectiveness are not straightforwardly linked. 
Then, after a performance evaluation aiming to clarify the difference between communication performance and goal-effectiveness, a long-term optimization problem is formulated and solved via Lyapunov stochastic network optimization tools, to guarantee the desired target performance. Finally, our numerical results assess the advantages of the proposed optimization, and the superiority of the goal-oriented strategy against baseline 5G compliant legacy approaches, under both stationary and non-stationary communication (and computation) environments. 
\end{abstract}



\maketitle

\section{Introduction}\label{sec:intro}
Today, as the fifth mobile generation (5G) is at its deployment stage\footnote{\url{https://www.3gpp.org/technologies/5g-system-overview}}, the race to 6G has started all around the world \cite{uusitalo2021, saad2020}. The latter entails theoretical research of new technologies' fundamental limits, the definition of new cutting edge use cases \cite{hexa-XD1.2,hexa-XD1.3}, their associated Key Performance Indicators (KPIs) and Key Value Indicators (KVIs) and, obviously, the new technological levers to enable the ecosystem. New technological enablers are required at all layers of the protocol stack, from application and network to the physical layer \cite{hexa-XD4.2}, but also down to the wireless propagation environment with reconfigurable intelligent surfaces \cite{Strinati_RIS_2021}. This cross-layer perspective will give birth to a fully reconfigurable system that can adapt to extremely diverse requirements, with a joint orchestration of heterogeneous resources (wireless, computing, storage). 
Following the revolution provided by Multi-access Edge Computing (MEC) \cite{kekki2018etsi}, a paradigm shift in 6G will be 
the integration of communication, computing and storage, natively enabling \textit{heterogeneous intelligence} (e.g., humans, robots, or machines) 
to accomplish complex cooperative tasks \cite{Letaief2022,Yang2020,Zhang2020,park2019wireless,Zhou19_EI}.
Connecting intelligence requires informative cooperative interactions. New trends in 6G explore both the \textit{semantic problem} of understanding the meaning of the source data and, the \textit{effectiveness problem}, whose aim is that of accomplishing target goals, through possibly corrupted/compressed/encoded/misunderstood received information \cite{Strinati_semantic2021}.

Achieving goals with target effectiveness requirements define a new family of KPIs in future 6G systems. We refer to such family of KPIs as \textit{goal-effectiveness}. This paper explore as main KPIs the \textit{goal-effectiveness} together with its associated \textit{goal cost}, which refers to the price to pay to achieve a goal.
Goal-effectiveness is one of the main KPIs to assess the performance of a \textit{goal-oriented (GO) communication} \cite{Goldreich2012,Strinati_semantic2021}, a true paradigm shift from \textit{data-oriented} (DO) communications \cite{Letaief2022}, envisioned to proliferate in 6G. While the aim of DO communications is for an intended data destination to reliably receive the original data generated by a data source, the aim of GO communications is rather for the data destination (playing the role of an acting agent) to receive adequate data of sufficient quality in the context of correctly performing a (possibly cooperative) task. Indeed, in our view, GO communications can refer to:
\begin{itemize}
    \item \textit{goal-oriented data compression} \cite{ZouZhang2021,binucci2022,ZhangZou2022,pezone2022, pappas2021,mostaani2019,Merluzzi_desiree,Merluzzi2022_ICC,MerluzziEML2021,Letaief2022,Shao2020,Xie2022}, aiming at extracting and adapting the relevant information needed to make the receiver accomplish a goal with desired effectiveness. This can be based on semantic information extraction \cite{pappas2021,Sana2022,Xie2022}, but it is not restricted to the meaning of data;
    \item \textit{goal-oriented transmission} \cite{Lee19, Merluzzi2022GO, Mu2022}, aiming at adapting communication reliability (e.g., the Packet Error Rate - PER), to achieve target goal-effectiveness, i.e., the probability of achieving the goal. This can also involve semantic-aware packet protection, under the assumption that some packets bring more relevant information than others, from the perspective of the goal.
\end{itemize}
Both concepts above also entail the definition of costs, which can include, among the others, energy consumption, delay, resource utilization, and communication overhead. Obviously, a combination of both approaches is possible and would represent a unified view towards a fully GO system design. 

As a second aspect tackled in this paper, a fundamental challenge of future 6G systems is to accommodate, on the same network infrastructure, services that are heterogeneous from several perspectives: requirements, resource consumption, time-variability, etc. Therefore, while it is true, in our view, that 6G will enable new classes of GO and semantic services, it is also sure that it will keep serving classical users and/or \textit{verticals} \cite{maman2021beyond} (i.e., data-oriented services), which in the sequel we refer to as \textit{legacy communication or, data-oriented} systems and services (in this paper, we will refer to them as \textit{DO systems} for brevity). For this reason, at some point in the near future (e.g., in the next $10$ years), a natural question will arise on the coexistence of GO and DO systems, to understand how one system affects the other, when they share the same network infrastructure and wireless resources. A possible question in this context is:\\ \indent\textit{How does the interference of a DO system affect GO performance, and at which cost?}\\
A similar question is considered in \cite{LEE2022}, where the authors analyze the performance of a system in which a semantic and a DO service coexist.

In this paper, we aim to answer the question of coexistence of GO and DO systems, based on a formal definition of a goal, entailing effectiveness and costs that jointly take into account GO and DO system requirements. After a general, yet, formal definition of goal and GO optimization problems, we will consider the case of MEC-enabled edge inference as a use case for the GO system. In the proposed setting, a GO system aims to classify data collected by an end device and transmitted to a nearby Mobile Edge Host (MEH), while a DO system aims to upload data with the highest possible data rate. The fundamental question to answer is then the following:\\
\indent\textit{What is the highest DO system data rate that can be achieved while, at the same time, guaranteeing a target goal-effectiveness of the GO system, assuming full spectrum sharing (i.e., in the presence of co-channel interference)?}

\subsection{Related work}\label{sec:literature}
Although the semantic and goal-oriented communication paradigm is in its infancy within the communication community, several works already started investigating the fundamental limits and the algorithmic foundations.
In \cite{kountouris2021}, the authors propose an end-to-end semantic communication model, entailing sampling, filtering, preprocessing, reception, and control.  In \cite{ZouZhang2021}, a GO data quantization scheme is proposed, to reduce the communication overhead of a decision-making service. Similarly, a GO data quantization and data clustering approach is proposed in \cite{ZhangZou2022}. In \cite{binucci2022}, the authors propose an online algorithm able to adapt to channel conditions, through the selection of a GO compression scheme in an edge inference scenario. Similarly, \cite{pezone2022} proposes a GO data compression method based on the information bottleneck principle. The authors of \cite{pappas2021} show the gain of a semantic-aware transmission on the reconstruction error of a real-time tracking system, while \cite{mostaani2019} focuses on the cooperation between agents with a common goal, through a task-oriented mutual information exchange. In \cite{Merluzzi_desiree}, it is shown how cooperation of different Machine Learning (ML) models can help improving the energy efficiency of an edge inference system with quantized data transmission, and a similar GO resource allocation based on an adaptive selection of JPEG compression is proposed in \cite{Merluzzi2022_ICC}. In \cite{Shao2020}, a goal-aware DNN splitting and feature extraction is proposed in an edge inference scenario, while in \cite{Xie2022}, a unified GO semantic communication framework is proposed in the context of a visual question answering use case. From a GO transmission perspective, the authors of \cite{Lee19} propose a joint transmission and recognition scheme, showing the effect of communication reliability on the accuracy, however not considering delay and energy consumption, as well as cooperative inference. Also, in our precursor conference paper \cite{Merluzzi2022GO}, we investigate the performance of communication KPIs (i.e., the bit error rate) on a real-time edge inference task, taking into account classification accuracy, energy, and delay. Furthermore, \cite{LEE2022} introduces, for the first time, the concept of coexistence between semantic and DO services, with a performance evaluation aimed to achieve sum-rate maximization of DO users, with a minimum required SNR constraint for semantic users, with mutual interference. However, the authors of \cite{LEE2022} do not consider goal-effectiveness constraints in the presence of interfering DO users, and do not propose a GO optimization, which is the purpose of this paper for an edge inference use case. 

Finally, \cite{Mu2022} and \cite{Mu2022_conf}, which are the most closely related works, exploit semantic communication for non-orthogonal multiple access (NOMA), by considering a system with a primary and a secondary user served by the same Access Point (AP) through NOMA. The authors show the gain of defining a semantic ergodic rate in the overall system performance, with the aim of maximizing the primary (i.e., bit-level communication) user data rate, under a semantic ergodic rate constraint of the secondary user, controlling transmit power and transmission time scheduling. Starting from the research question of \cite{Mu2022} and \cite{Mu2022_conf}, we investigate a novel scenario, in which a GO and a DO communication system coexist, thus focusing on GO, rather than semantic, communication.

\subsection{Our contribution}\label{sec:contributions}
Similarly as \cite{Mu2022} and \cite{Mu2022_conf}, in this paper, we aim at analyzing the impact of mutual interference on future service classes. Differently from \cite{Mu2022} and \cite{Mu2022_conf}, whose focus is on semantic communication and ergodic (semantic) rate, and where a single base station serves users with bit and semantic streams, we rather focus on goal-oriented communication and resource allocation in connect-compute networks, in which a goal-oriented (GO) and a data-oriented communication (DO) system coexist. Therefore, the main contributions of our work are the following:
\begin{itemize}
    \item We consider the communication effectiveness in accomplishing a predefined goal/task, as our system constraint. Therefore, the focus of our work is on GO communications (i.e., effectiveness of communication toward achieving a goal - and in particular through resource allocation) rather than semantic communications (i.e., understanding the meaning of data).
    \item We introduce the edge inference service as the main use case under investigation, defining it as a GO communication service and problem.
    \item We consider the communication reliability (i.e., Packet Error Rate - PER) as a variable to be controlled to achieve target goal-effectiveness at the GO system, while maximizing performance of the DO system, in terms of data rate.
    \item We propose a \textit{computation resource-aware} method for guaranteeing goal-effectiveness, taking into account the computing resource availability for edge inference.
\end{itemize}
In other words, our work is about goal-aware system coexistence, while \cite{Mu2022} and \cite{Mu2022_conf} are about a multiple access scheme where different traffic types (semantic, bit-oriented) are multiplexed by a single AP that can process both sorts of data. 

Then, as a first step of this work, we formally define a GO resource optimization problem in the context of wireless communication systems. This will help us introducing, in the sequel, the fundamental problem of coexistence of GO and DO systems in 6G, with full spectrum sharing. As a specific use case, we consider an edge inference service for the GO system, in which an end device uploads data to a MEH, which runs a classification task through a pre-trained and pre-uploaded ML model (here, a Convolutional Neural Network - CNN). In such a system, we will define goal-effectiveness and goal cost, with the former relating to the probability of receiving confident classification results within a predefined deadline, and the latter relating to the data rate loss of the DO user, compared to the case in which the GO user is absent. This bonds interference, communication reliability, computing resources, goal-effectiveness, and goal cost into a unified framework. To the best of our knowledge, this has never been done before in the literature for GO communication and edge inference scenarios, jointly factoring in wireless and compute resources. 

After a performance evaluation of the proposed scenario, aiming to gain insights about the system behaviour as a function of different free system parameters, we will propose a simple, yet, relevant optimization problem, along with an adaptive algorithm able to optimize radio resources to minimize the goal cost, under goal-effectiveness constraints, with a connect-compute resources-aware approach. In summary, the main novelty of the paper consists in conducting a GO performance analysis of an edge inference system sharing resources with a DO system, along with an adaptive algorithm to attain desired performance, being aware of application constraints, its online performance, as well as connect-compute resource availability. After proposing an algorithm with theoretical guarantees, we test it on scenarios with different parameters and requirements, against baseline strategies (e.g., orthogonal bandwidth splitting), but also on non-stationary environments in which GO requirements (e.g., goal-effectiveness) or computation resource availability statistics, can unexpectedly change over time.
\subsection{Organization of the paper}\label{sec:paper_org}
The remainder of this paper is organized as follows: Section \ref{sec:goal_definition} provides a general definition of a goal in the context of wireless communications, entailing effectiveness, constraints, costs, and goal achievability; Section \ref{sec:multi_user} focuses on applying the definition to the proposed scenario, in which GO and DO systems coexist, fitting the general definitions to the specific use case of edge inference (GO user) and data upload (e.g., video streaming - DO system). Also, a numerical evaluation without optimization is provided, to explore performance and gain insights about the relevant parameters that will be considered as optimization variables in Section \ref{sec:problem_formulation}, in which a long-term problem formulation, along with a solution based on Lyapunov stochastic network optimization, is proposed. Numerical results assessing the performance of the proposed algorithm are presented in Section \ref{sec:numer_eval}, while Section \ref{sec:conclusions} draws the conclusions of the paper, and proposes some future research directions. 
\subsection{Notation and acronyms}
\begin{table}[t]
\caption{List of acronyms}
\label{tab:my-table}
\resizebox{\columnwidth}{!}{%
\begin{tabular}{|l|l|l|l|}
\hline
AI   & Artificial Intelligence      & GPU  & Graphical Processing Unit                  \\ \hline
AP   & Access Point                 & KPI  & Key Performance Indicator                  \\ \hline
CLD  & Conditional Lyapunov Drift   & KVI  & Key Value Indicator                        \\ \hline
CNN  & Convolutional Neural Network & MEC  & Mutli-access Edge Computing                \\ \hline
CPU  & Central Processing Unit      & MEH  & Mobile Edge Host                           \\ \hline
DNN  & Deep Neural Network          & MCS  & Modulation and Coding Scheme               \\ \hline
DO   & Data-Oriented                & ML   & Machine Learning                           \\ \hline
DPP  & Drift Plus Penalty           & NREI  & Negative relative average entropy increase \\ \hline
DRL  & Deep Reinforcement Learning  & NOMA & Non-Orthogonal Multiple Access             \\ \hline
E2E  & End-to-End                   & PER  & Packet Error Rate                          \\ \hline
eMBB & enhanced Mobile Broad Band   & SINR & Signal-to-Interference-plus-Noise Ratio              \\ \hline
GO   & Goal-Oriented                & UE   & User Equipment                             \\ \hline
\end{tabular}%
}
\end{table}
Throughout the paper, bold lower case and upper case letters denote vectors and matrices, respectively; the operator $|\cdot|$ is the absolute value of a complex number, and the superscript $^H$ denotes the Hermitian operator; $card(\cdot)$ denotes the cardinality of a set, and $\mathbf{1}_{A}$ denotes the indicator function, which equals $1$ if condition $A$ is satisfied, and $0$ otherwise. Finally, calligraphic upper case letters always denote the long-term average of a corresponding variable throughout the text; for instance, for variable $x$ (or $X$), the long-term average is denoted as $\mathcal{X}$, and it reads as follows:
\begin{equation}\label{long_term}
    \mathcal{X}=\lim_{T\to\infty}\frac{1}{T}\sum_{t=0}^{T-1}\mathbb{E}\{x_t\},
\end{equation}
where $t$ indicates a temporal index. The acronyms used throughout the paper are define the first time they appear, but can be also found in Table \ref{tab:my-table}, in alphabetic order.

\section{Definition of a goal}\label{sec:goal_definition}
The goal-oriented communication paradigm constitutes a communication approach for which performance is not measured by classical metrics, such as data rate or wireless link reliability, but rather by the success level with which a network entity is able to perform a sequence of application-related tasks, as a result of exchanging data with one or more other network entities. In the sequel, we will also refer to these entities as \textit{agents}. In this context, an agent is any entity endowed with communication capabilities, which could also have computing capabilities and be possibly embarked with Artificial Intelligence (AI) and/or ML models. In the latter case, we can refer to it as \textit{AI agent}. However, this distinction is not essential in this paper.

A fundamental step is to formally define a goal, to be able to properly formulate goal-oriented communication problems. Indeed, a wrong, inexact, unclear, or ambiguous definition of a goal, may lead to dramatically wrong or biased decisions and communication/computation policies. \\
In \cite{Hollnagel91}, E. Hollnagel defines a goal out of the context of wireless communications. 
From its definitions, and with the aim of embedding the goal-oriented perspective into wireless communication systems, in the most general case, a goal is characterized by a set of requirements that, if guaranteed, determine its accomplishment. 
When embedded in wireless scenarios with inter-agent communications, the accomplishment of these requirements is directly linked to communication (and computation) related performance and strategies. As an example, suppose that the goal of the communication is to exchange sampled data for anomaly detection. Then, data corrupted by a noisy (and possible interference-prone) channel could lead to wrong decisions, depending on the wireless link reliability (e.g., the PER). At the same time, data encoding schemes (e.g., compression) affect the goal achievement due to distortion. Both data encoding and wireless link reliability play a key role in GO communications. A priori, it is not generally easy to predict the exact correlation between goal accomplishment and wireless communication performance, data encoding, etc. This depends on several aspects, including the goal definition, the connect-compute network conditions, and the a priori knowledge of communicating agents.
Learning and timely adapting communication parameters towards goal accomplishment is the main target of the GO communication paradigm. 
As also remarked in \cite{Hollnagel91}, a goal is achieved through a series of tasks, with a task being a collection of actions. Generally, an action entails a decision on a set of parameters, which could involve communication, computation, and possibly control. Let us formally define the three measures that characterize a goal, and are necessary to formulate GO communication problems.

\subsection{The goal value, achievability, and effectiveness}
As anticipated, the first fundamental step is for an agent to be able to verify whether a goal has been achieved, by carefully selecting metrics assessed by the agent for that purpose. In this context, we define a quantified representation of the target system state indicating goal accomplishment, which we name after as \textit{goal value}. The goal value can be, for instance, a binary variable that equals $1$ if the goal is achieved, and $0$ otherwise. Of course, it is not restricted to this case, and it may be represented as a generic real vector (e.g., the position in space of a robot with respect to a target one). In a dynamic system, a goal value is possibly time-varying, i.e., at each time instant $t$, a new/updated goal value can be made available to an agent, as a result of environment states, agents' actions, and environment feedback. Therefore, in the most general case, the goal value can be denoted as a real variable $\Theta_t\in\mathbb{R}^{l}$, with $t$ denoting the time instant. 

\subsubsection{Goal-achievability and goal-effectiveness }\label{sec:goal_effect}
Going beyond the goal value, and, with focus on either repetitive or long-standing application-related goals, agents eventually have to measure the goal accomplishment rate, i.e., the probability of achieving the goal value. Indeed, a goal accomplishment must not necessarily be deterministic. For instance, a goal value can be obtained with a certain probability, which we can refer to as goal-effectiveness. The goal-effectiveness is inherently non negative and is a long-term measure. Indeed, even in the case a goal is a sequence of sub-goals, the sub-goal accomplishment is stochastic, and what we are interested in is the success rate of the goal. We can give the following example: \textit{i)} Edge pattern (e.g., image) classification: in this use case, we can choose the goal value to be equal to $1$ if a pattern is correctly classified, and $0$ otherwise. In this case, without additional constraints, the goal-effectiveness is the probability that the goal value equals $1$, and corresponds to the correct classification rate.

Nonetheless, besides the goal value itself, the goal-effectiveness may be subject to $J$ instantaneous constraints that, if not met, prevent the goal from being perceived by the agent as accomplished, per a user service level agreement. Sticking to the edge classification use case, the goal value is defined as before, while a possible constraint is represented by the End-to-End (E2E) delay being kept under a predefined threshold. Without loss of generality, we can write these constraints (which we refer to as \textit{short-term goal constraints}) as $g_{j,t}\leq 0, j=1,\ldots, J$. Building on the goal value and the short-term goal constraints, we can formally define the goal-effectiveness as follows: 
\begin{equation}\label{goal_effectiveness}
    \mathcal{E}_g =
    \lim_{T\to\infty}\frac{1}{T}\sum_{t=0}^{T-1}\mathbb{E}\left\{\mathbf{1}_{\{\Theta_t\geq \Theta_{\textrm{th}}\}}\cdot\prod_{j=1}^J\mathbf{1}_{\{g_{j,t}\leq 0\}}\right\},
\end{equation}
where $\Theta_{\textrm{th}}$ is a predefined goal value threshold, and the expectation is generally taken with respect to the random context parameters and the strategy (policy) followed by the agents to achieve the goal. From \eqref{goal_effectiveness}, we can already state that, as per our definition, the goal-effectiveness is a probability, thus $\mathcal{E}_g\in[0,1]$.

Going further, besides the goal value, achieving a goal may also entail other $M$ long-term constraints. As an example, a typical long-term constraint is to keep the average energy consumption of an agent below a predefined threshold, e.g., to not quickly drain its battery level. 
In the most general case, we can write these long-term constraints as follows:
\begin{equation}\label{long_term_constraints}
    \mathcal{G}_s:=\lim_{T\to\infty}\frac{1}{T}\sum_{t=0}^{T-1}\mathbb{E}\left\{G_{s,t}\right\}\leq \mathcal{G}_{s,\textrm{th}},\quad\forall s=1,\ldots,S.
\end{equation}

Building on \eqref{goal_effectiveness} and \eqref{long_term_constraints}, we can write the following definition for goal achievability:
\textit{i)} a goal is \textit{achievable} if there exists at least one policy that guarantees \eqref{goal_effectiveness} to be above a predefined threshold $\mathcal{E}_{g,\textrm{th}}$; \textit{ii)} a goal is \textit{strongly achievable} if there exists at least one policy that guarantees \eqref{goal_effectiveness} to be above a predefined threshold $\mathcal{E}_{g,\textrm{th}}$, while guaranteeing all other long-term constraints \eqref{long_term_constraints}. 
In typical communication system management, long-term constraints entail communication KPIs, e.g., PER below a threshold. This translates into, e.g., adaptive Modulation and Coding Scheme (MCS) selection, able to guarantee the target PER. Instead, with the goal-oriented paradigm, \emph{communication} KPIs are not necessarily required to be satisfied a priori, while measures belonging to the physical/human world, or, in general, to the application, are used to assess the performance of the communication system. In this case, classical communication KPIs can be \textit{learnt from experience}, and \textit{dynamically adapted}, to achieve the target goal-effectiveness in \eqref{goal_effectiveness} in the most efficient way, possibly subject to constraints \eqref{long_term_constraints}. 

\subsection{The goal cost}\label{sec:goal_cost}
Every goal should be accompanied by a cost spent to achieve it, i.e., a price to pay in terms of, e.g., radio and computation resources. We refer to this price as \textit{goal cost}. Indeed, there are possibly several strategies achieving the goal with target goal-effectiveness, i.e., a goal is possibly achievable via multiple strategies. However, different strategies have also different costs. The aim of goal-oriented communications is to find the strategy that achieves the target goal-effectiveness with the lowest goal cost. We can differentiate between short-term (or even instantaneous) and long-term costs. In the case of a long-term definition of the cost, we can define an average or a cumulative cost, with the latter being the sum of all instantaneous costs spent to achieve the goal. Normally, the latter refers to a strategy to achieve a goal, where, goal success is indicated by reaching a specific terminal state for the (physical) system, e.g., for a robot going from point A to point B, i.e., an \emph{episodic task}, as defined in (Deep) Reinforcement Learning (DRL). On the other hand, the average cost can also be used for those goals that last virtually infinite time, e.g., continuous collection and classification of patterns. Denoting by $C_t$ the instantaneous cost, we can define a long-term cost as follows (cf. \eqref{long_term}):
\begin{align}\label{goal_cost_avg}
    &\mathcal{C}=\lim_{T\to\infty}\frac{1}{T}\sum_{t=0}^{T-1}\mathbb{E}\left\{C_t\right\}.
\end{align}
In what follows, we aim to shed lights on how communication service requirements may impact goal-effectiveness and propose formulating the overall GO communication problem involving selection of policies across communication, computation and control domains.

\subsection{Identifying goal-achieving communication KPIs}\label{sec:general_problem_form}
In edge AI/ML scenarios and use cases, multiple agents (at least 2), equipped with more or less powerful computing units and complex (ML) models, communicate to achieve a (possibly common) goal. The issue is then how to allocate connect-compute resources in order to guarantee a target goal-effectiveness subject to short and long-term constraints, paying the lowest possible goal cost. First, when exchanging data, availability of communication resources affects the level of distortion on the received data with respect to the original information. In classical communication systems, communication KPIs are defined \textit{a priori}, depending on the different service requirements. This paradigm has started a long race towards adding "$9$s" to the communication reliability \cite{popovski14} in terms of e.g., packet success rate. Here, based on the definitions provided in the previous section, we argue that this is not the most efficient way of designing the system and orchestrating operations and resources. Indeed, we do not formulate a problem in which communication KPIs are explicitly taken into account, but they are rather controlled by learning and adaptation to communication policies that achieve target levels of goal-effectiveness. In this context, let us denote by $\pi_{\textrm{comm}}$ a \textit{communication policy} entailing, in the most general case: \textit{i)} source encoding (e.g., data compression schemes) \cite{MerluzziEML2021,Merluzzi2022_ICC}; \textit{ii)} modulation and channel coding \cite{MerluzziDMEC}; \textit{iii)} wireless channel scheduling, including node participation selection \cite{BattiloroFL2022} and association \cite{Sana19} ; \textit{iv)} multiple antenna transmission scheme, devising power allocation and precoding \cite{Hao17}. Moreover, let us denote by $\pi_{\textrm{comp}}$ a \textit{computation policy} entailing, in the most general case: \textit{i)} local computing resource scheduling at each device/agent; \textit{ii)} computation resource scheduling at shared computing units (e.g., in MEHs) \cite{Mao2016,Mao2017,Merluzzi2020URLLC}; \textit{iii)} selection of a single one or multiple collaborative AI agents containing relevant ML models for ensemble-based inferencing  \cite{Merluzzi_desiree,Shlezinger2021} or federated learning \cite{BattiloroFL2022}. Finally, we denote by $\pi_{\textrm{contr}}$ a \textit{control policy} entailing AI agents mobility and trajectories. 

A general GO communication problem can be then formulated as follows:
\begin{align}\label{general_prob_form}
    \underset{\{\pi_{\textrm{comm}},\pi_{\textrm{comp}},\pi_{\textrm{contr}}\}}{\min}\;&\mathcal{C}:=\lim_{T\to\infty}\frac{1}{T}\sum_{t=0}^{T-1}\mathbb{E}\left\{C_t\right\}\\
    &\text{subject to}\nonumber\\
    &(a)\;\pi_{\textrm{comm}}\in\mathcal{P}_{\textrm{comm}}\nonumber\\&(b)\;\pi_{\textrm{comp}}\in\mathcal{P}_{\textrm{comp}}\nonumber\\
    &(c)\;\pi_{\textrm{contr}}\in\mathcal{P}_{\textrm{contr}}\nonumber\\
    &(d)\;\mathcal{E}_g\geq \mathcal{E}_{g,\textrm{th}}\nonumber\\
    &(e)\;\mathcal{G}_s\leq \mathcal{G}_{g,\textrm{th}},\;\; s=1,\ldots,S\nonumber,
\end{align}
where $\mathcal{P}_{\textrm{comm}}$, $\mathcal{P}_{\textrm{comp}}$, $\mathcal{P}_{\textrm{contr}}$ denote the feasible sets of communication, computation and control actions, respectively. The constraints in \eqref{general_prob_form} have the following meaning: $(a)$ the communication policy belongs to the feasible set $\mathcal{P}_{\textrm{comm}}$ (e.g., the transmit power is non negative and lower than a threshold); $(b)$ the computation policy belongs to the feasible set $\mathcal{P}_{\textrm{comp}}$ (e.g., the sum of all CPU resources of an MEH assigned to a set of users does not exceed the maximum CPU computing power); $(c)$ the control policy belongs to the feasible set $\mathcal{P}_{\textrm{contr}}$ (e.g., a robot can move towards a predefined set of possible directions); $(d)$ the goal-effectiveness is above a predefined threshold (e.g., the inference accuracy of an edge inference system is above a threshold); $(e)$ other long-term constraints are satisfied (e.g., the average energy of an agent does not exceed a predefined threshold). The problem formulation in \eqref{general_prob_form} is quite general, and it can be easily customized to specific use cases. In the next section, we present a specific scenario in which a GO and a DO communication system coexist. The aim is to hinge on the GO paradigm and formulation, to maximize the DO system data rate under goal-effectiveness constraints of the GO system.

\section{Coexistence of a goal-oriented and a data-oriented communication system}\label{sec:multi_user}
\begin{figure*}[t]
    \centering
    \includegraphics[width=\textwidth]{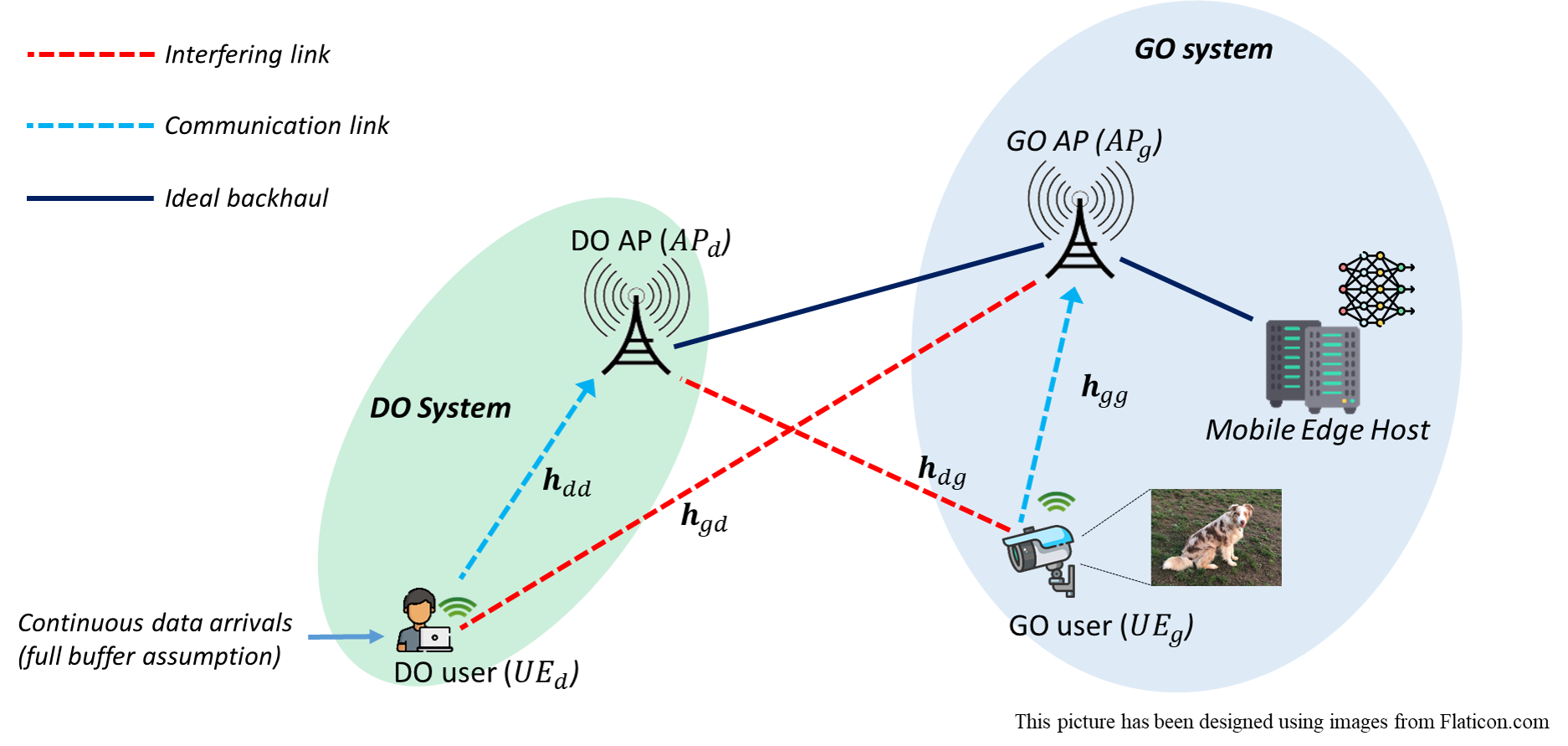}
    \caption{Reference scenario: a GO edge inference system coexists with a legacy  DO system; the two systems generate mutual interference, which affects inference performance (goal-effectiveness) and DO user data rate (goal cost).}
    \label{fig:CR_GO}
\end{figure*}
In this section, we apply the GO approach to a scenario in which two different systems coexist: \textit{i)} a GO communication system, consisting of a User Equipment (UE) offloading inference tasks to an MEH collocated to an AP, sharing radio spectrum with \textit{ii)} a classical DO communication system, consisting of another UE connected to another AP, where, the device continuously uploads content, such as a video stream, to be cached in the network. As an example, the focused scenario can refer to an industrial setting for low-power networking, where, a device uploading product images for anomaly detection (GO transmitter/ receiver pair) coexists with a security camera providing continuous video feed of the factory floor (legacy DO eMBB service). We believe that such a scenario will be of extreme interest in the future, as in beyond 5G and 6G systems, the same system architecture will need to support both such services. As spectrum resources under 6 GHz will continue to be pivotal in providing wide radio coverage \cite{Hexa-X2021}, (non-orthogonal) spectrum sharing is envisioned to be a standing feature, especially for low-power networking, e.g., in an industrial environment.

The GO approach may find direct application to system scenarios involving coexistence of services of different types. For such services, the most relevant and important communication KPIs (e.g., data rate, latency, reliability) may even be the same, however, the tasks performed may substantially differ in scope, hence design trade-offs need to be addressed to concurrently guarantee all application-related goal values by keeping goal costs as low as possible for all tasks. To further elaborate on the problem of \emph{conflicting goals}, in this paper we choose to investigate the coexistence of the two systems, inspired by the Cognitive Radio (CR) networking paradigm \cite{mitola1999cognitive, haykin2005cognitive}. The fundamental difference lies in that, for the CR problem, where a primary and a secondary system operate together in space and time, focus is, in general, on allocating the available secondary radio resources to maximize secondary system performance, subject to the non-violation of \emph{radio} performance constraints imposed by the primary system operator. Nonetheless, for the focused GO/DO system coexistence, the ultimate design target is \emph{service-effective communication} even with ``softened'' radio communication requirements for the GO system that can be learned during goal accomplishment.

\subsection{System setup}\label{sec:CRN_setup}
The system of interest is illustrated in Fig.~\ref{fig:CR_GO}. A GO transmitter-receiver pair (which we, in short, term after as GO system), consisting of a single-antenna UE, denoted as $UE_g$ and referred to as \textit{GO user}, and an AP, denoted as $AP_g$ and referred to as \textit{GO AP}, equipped with $M_g$ antennas fully shares the available radio frequency spectrum of bandwidth $W$ (Hz) with a DO transmitter-receiver pair (which we, in short, term after as DO system) of the same characteristics, i.e., a single-antenna UE, denoted as $UE_d$ and referred to as \textit{DO user} and an AP, denoted as $AP_d$ and referred to as \textit{DO AP}, equipped with $M_d$ antenna elements. It is assumed that $AP_g$ and $AP_d$ are interconnected via a backhaul link characterized by high capacity and low delay. Besides radio infrastructure, it is assumed that an MEH is collocated with $AP_g$, where, the interconnection delay can be considered as negligible. Generalizing this would be straightforward, and is left for future investigations.

Uplink data communication takes place for both systems during operation time. For the GO system, uplink data communication is carried out to offload a batch of data patterns (e.g., images, either received or generated at $UE_g$) to the MEH, to obtain classification information as output, by means of a downlink transmission (whose delay is considered negligible in this work). In contrast to the GO system, the aim of DO system communication is for $UE_d$ to provide an incoming data stream, e.g., a video feed, to $AP_d$. Storage and distribution of this content are not investigated in this paper and are left for future investigations.

At the physical layer, we consider an uplink Single-Input Multiple-Output (SIMO) system. We denote by $\mathbf{h}_{ij,t}\in\mathbb{C}^{M\times 1}$, the uplink channel vector between $UE_j$ (with $j=g,d$) and $AP_i$ (with $i=g,d$) at time slot $t$, which can be written as 
\begin{equation}\label{channel_matrix}
    \mathbf{h}_{ij,t} = \frac{\sqrt{\beta_{ij,t}^{-1}}}{\sqrt{(K+1)}}\left(\sqrt{K}\mathbf{h}_{ij,t,\textrm{LOS}}+\mathbf{h}_{ij,t,\textrm{NLOS}}\right),
\end{equation}
where the $t$-th component of the Line-Of-Sight (LOS) component vector reads as $\mathbf{h}_{ij,\textrm{LOS}}(t)=e^{-\frac{j2\pi d_{mij,t}}{\lambda}}$, with $\lambda$ the wavelength, $d_{mij,t}$ the distance from $UE_j$ single antenna to the $m$-th antenna of $AP_i$, while the Non-LOS (NLOS) component follows a circularly symmetric Gaussian distribution, i.e., $\mathbf{h}_{ij,t,\textrm{NLOS}}\sim \mathcal{C}\mathcal{N}(0,1)$. Path loss between $UE_j$ and $AP_i$ is represented by $\beta_{ij,t}$, while $K$ is the Rician factor, strongly dependent on the scenario (e.g., indoor, outdoor, etc.). 

In this reference setting, both system communication and computing resources are involved. From a computation point of view, we assume an MEH hosting a relevant ML model (e.g., a Deep Neural Network - DNN, that is assumed to be already trained), which dynamically assigns resources to the GO system, based on its current availability. Therefore, on a per-slot basis, the MEH communicates with $AP_g$ to select an inferencing pattern offloading policy, based on current connect-compute conditions, and based on the goal cost and goal-effectiveness definition. 

\subsection{Edge inference: goal value and effectiveness}
In what follows in this section, we will define goal value, effectiveness, and cost for the focused system. To do so, let us first consider the communication, computation, and inference performance of the GO system. Indeed, as already mentioned, and as it will be clarified for this particular scenario, we are interested in assessing system performance in terms of inference delay, entailing communication and computing phases, and inference confidence, which translates into inference accuracy. These two measures, when jointly considered, define the goal-effectiveness for edge inference, as it will be formalized later on in this section.

\subsubsection{Uplink radio performance of GO user}
Let us recall that the GO user offloads inference tasks to the MEH. From a wireless perspective, performance is affected by multiple factors: \textit{i)} the uplink data rate; \textit{ii)} the interference received from the DO user; \textit{iii)} the target packet error rate. Let us denote by $\mathbf{w}_g\in \mathbb{C}^{M_g\times 1}$ the combining vector at $AP_g$, denoting by $P_g$ the transmit power of $UE_g$, and by $P_d$ the transmit power of $UE_d$. Assuming time as organized in slots $t=1,2,3,\ldots$, the Signal-to-Interference-plus-Noise Ratio (SINR) at the GO receiver ($AP_g$), at time $t$, reads as
\begin{equation}\label{SINR_p}
    \textrm{SINR}_{g,t}=\frac{|\mathbf{w}_{g,t}^H\mathbf{h}_{gg,t}|^2P_{g,t}}{N_0W+|\mathbf{w}_{g,t}^H\mathbf{h}_{gd,t}|^2P_{d,t}},
\end{equation}
where index $t$ denotes the time dependence of the involved random and controlled variables; whereas, $N_0$ denotes the power spectral density, and $W$ the uplink bandwidth, assumed to be fully shared between the two systems. Now, considering finite blocklength transmissions \cite{Polyanskiy2010,Sun19,She17}, and denoting by $\gamma_t$ the target PER at time $t$, we can write the achievable rate of $UE_g$ (in bits/s) at time $t$ as follows \cite{Polyanskiy2010}:
\begin{equation}\label{data_rate}
    R_{g,t}\! = \!W\!\left[\log_2\left(1+\textrm{SINR}_{g,t}\right)\!-\!\frac{1}{\log(2)}\sqrt{\frac{V_t}{n_g}}Q^{-1}(\gamma_t)\right],
\end{equation}
where $n_g$ is the blocklength, $Q^{-1}(\cdot)$ is the inverse of the Gaussian Q function, while $V_t$ is the channel dispersion, given by \cite{Sun19,She17}
\begin{equation}
    V_t=1-\frac{1}{\left(1+\textrm{SINR}_{g,t}\right)^2},
\end{equation}
which is well approximated as $V\approx 1$ with SINR above $5$ dB \cite{Sun19}. Nevertheless, in this work, we keep it general, as our system may be required to work at low SINR regimes, provided that a target goal-effectiveness is guaranteed. Finally, assuming (without loss of generality) that a batch of $N_t$ new patterns is requested to be inferred at time slot $t$, the uplink transmission delay for GO communication reads as follows:
\begin{equation}\label{delay_comm}
    D_{\textrm{tx},t}=\frac{N_t n_b}{R_{g,t}},
\end{equation}
where $n_b$ denotes the number of bits encoding one pattern. In this work, we do not focus on the specific encoding scheme, which we assume to be fixed. Semantic and GO compression can also play a role in optimizing $n_b$, which goes beyond the scope of this paper and is left for future investigations. Preliminary works on online compression level selection are available in \cite{MerluzziEML2021,Merluzzi2022_ICC,Merluzzi_desiree,Letaief2022,Shao2020,Lee19}.
\subsubsection{Computation delay of GO user}\label{sec:comp_delay}
Uplink transmission is just the first of the (at least two) phases of wireless edge inference, whose second step is remote processing through an ML model (e.g., a DNN), running in an MEH. The processing phase constitutes another source of delay, which is, typically, non-negligible, especially for demanding inferencing tasks served by MEHs of limited computing capabilities, as compared to the ones at the distant cloud. The computation delay is not, in general, a deterministic quantity, as it depends on CPU loading, execution of background processes, access to memory, etc. Obviously, it strongly depends on the employed ML model \cite{Canziani16}. In this paper, we assume that the MEH, irrespective of wireless and computing resource availability, always employs the same model (assumed already trained), and we leave GO model selection criteria for future investigations. As such, we denote by $D_{\textrm{comp},t}$, a random variable (whose statistics are possibly unknown), representing the computation time at the MEH. At each time slot, the computing time is possibly different, and its estimation is assumed to be available at the MEH, making it able to optimize connect-compute resources. 

Finally, the total inference time, neglecting downlink transmission of small size results (typically a few bits for binary or multi-class classification), is given by
\begin{equation}\label{delay_tot}
    D_{\textrm{tot},t}=D_{\textrm{tx},t}+D_{\textrm{comp},t}.
\end{equation}
Part of the goal is to retrieve inference results earlier than a predefined deadline $D_{\max}$, as clarified in the remainder of this section. However, retrieving incorrect inference results within the deadline can be extremely harmful for a special purpose functionality. For this reason, we need to define a measure of inference confidence, to be incorporated into the overall definition of goal-effectiveness.
\subsubsection{On the use of entropy to define the goal value}
We now introduce the metric reflecting inference accuracy, which will be then used to define the goal value. As we are dealing with a classification task, the actual performance metric could be the one of correct pattern classification rate. However, to evaluate this metric, the ground truth, i.e., the true label, would need to be always available, which is not the case, in general. In a practical setting, a metric suitable for evaluating classification tasks as they emerge, needs to be characterized by the following features: \textit{i)} it must be measurable online without any ground truth; \textit{ii)} it should reflect, in the best possible way, the value of the goal, i.e., the classification accuracy. 
A possible metric fulfilling these requirements is the entropy computed on a posteriori probabilities, which are the typical output of any discriminative or generative classifier, such as a DNN \cite{bishop2006pattern,Merluzzi2022_ICC,Saerens2002}. In particular, the output of a discriminative classifier can be written as a vector $\mathbf{p}=[p_1,\ldots,p_L]^H$ of probabilities, with each probability being associated to one out of the $L$ possible labels of the data set under investigation. Obviously $\sum_{l=1}^Lp_l=1$ holds true. Given $\mathbf{p}$, the entropy associated to a classified pattern $b$ is
\begin{equation}\label{entropy}
    H_b=-\mathbf{p}^H\log(\mathbf{p}).
\end{equation}
The entropy is a scalar measure, which can be interpreted as the classification confidence of an ML model, in classifying a pattern. The lower the entropy, the more confident the classifier is. Namely, for a data set with $L$ labels, the worst case for a single pattern classification is $p_l=\frac{1}{L}, \forall l=1\ldots,L$ (i.e., throwing an $L$ facets dice), which translates into the maximum entropy $H_{\max}=\log(L)$; also, noting that $\lim_{p\to 0^+}p\log(p)=0$, and that $p\log(p)=0$ if $p=1$, the minimum value attained by the entropy in case of a completely sure classifier is $0$. Therefore, the entropy is always a bounded metric $0\leq H\leq H_{\max}=\log(L)$. 
Since we are considering inference on batches of data (cf. \eqref{delay_comm}), we can define the \emph{average entropy} $H_t$ on the batch classified during slot $t$ as
\begin{equation}\label{batch_entropy}
    H_t=\frac{1}{N_t}\sum\nolimits_{b\in\mathcal{B}_t}H_{b,t},
\end{equation}
where $\mathcal{B}_t$ denotes the set of patterns in the batch, and $N_t=card(\mathcal{B}_t)$ (cf. \eqref{delay_comm}).
For simplicity, and, for the sake of coherence with the goal-value definition in Section \ref{sec:goal_definition}, we will use, as goal value, the negative relative average entropy increase (referred to as \textit{NREI} in the sequel), with respect to the its minimum value, attained on a validation subset of the original data set, i.e., without any communication impairment (interference from DO user)\footnote{It can be, for instance, assumed that model training is performed using a radio bandwidth (of same size as in inference phase) exclusively used by the GO system.}. This choice is dictated by the fact that, even for the original data set, the average entropy is not actually $0$, but it attains a minimum value that depends on the training phase. Therefore, denoting by $H_{\min}$ the minimum average entropy over the considered data set, we write the NREI (i.e., the goal value) as follows:
\begin{equation}\label{goal_value_EI}
    \Theta_t=-\frac{H_t-H_{\min}}{H_{\min}}.
\end{equation}
Then, the higher the absolute value of the NREI, the lower the goal value is, and vice versa. A natural question arises on the dependence of this goal value on wireless performance, e.g., on the PER. In this paper, since we consider the PER as source of classification entropy increase/decrease, we make the following principled and mild assumptions on entropy and accuracy, which we validate by experimental results.\\
\underline{\textit{Assumption 1:}} \textit{The classification entropy is a monotonic non-decreasing function of the PER.}\\
This assumption, despite not being supported by a mathematical proof, is based on the rationale that a more severely degraded version of the inferencing input data (i.e., distorted input patterns received by $AP_g$) generates higher ``confusion'' in the classifier, thus increasing the classification entropy. \\
\underline{\textit{Assumption 2:}} \textit{The classification accuracy is a monotonic non-increasing function of the average entropy.}\\
\begin{figure}[htb!]
    \centering
    \subfloat[Average entropy vs. PER]{
        \includegraphics[width=\columnwidth]{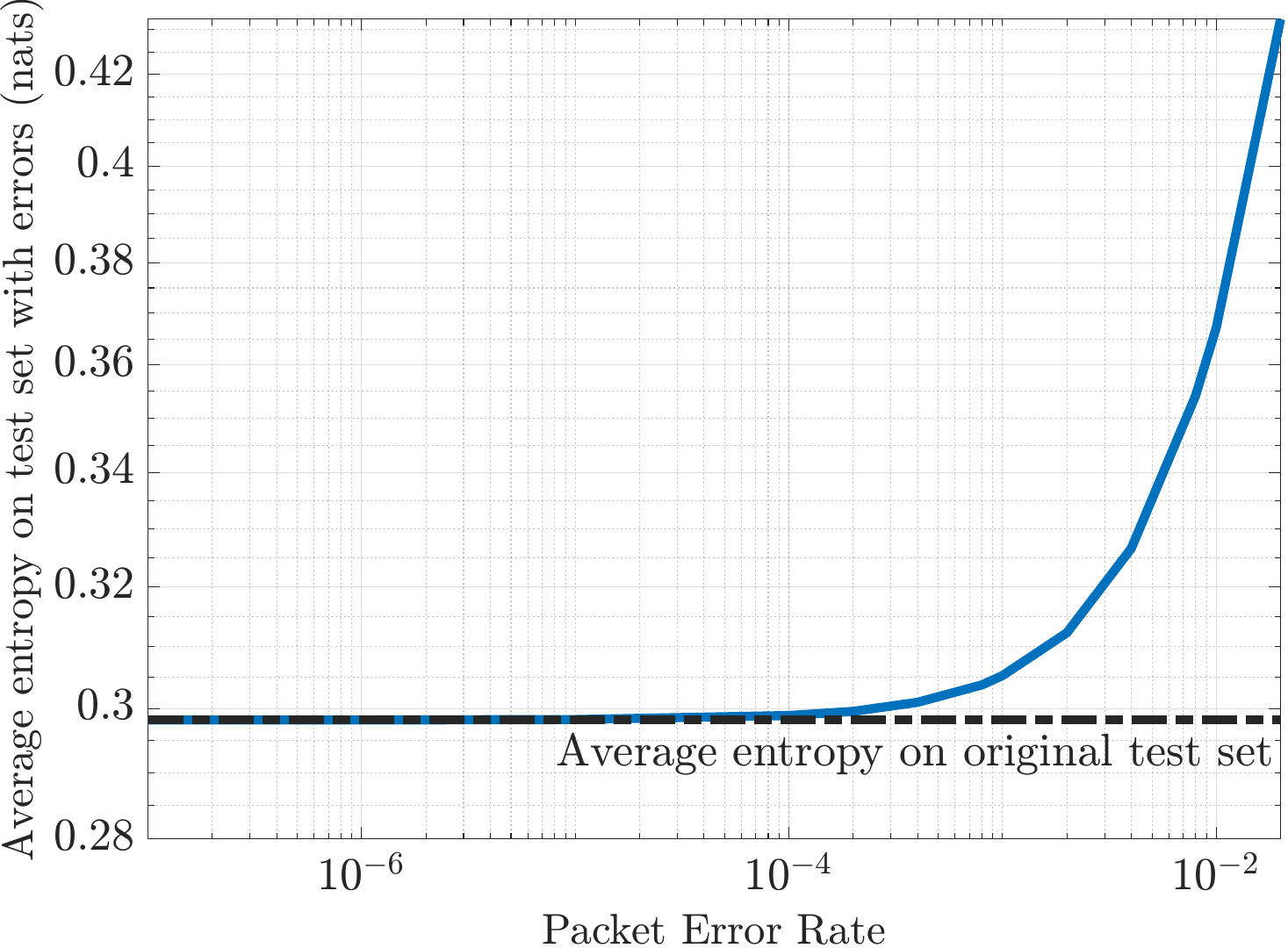}
        \label{fig:entropy_vs_PER}
    }
    
    \subfloat[Accuracy vs. entropy]{
        \includegraphics[width=.98\columnwidth]{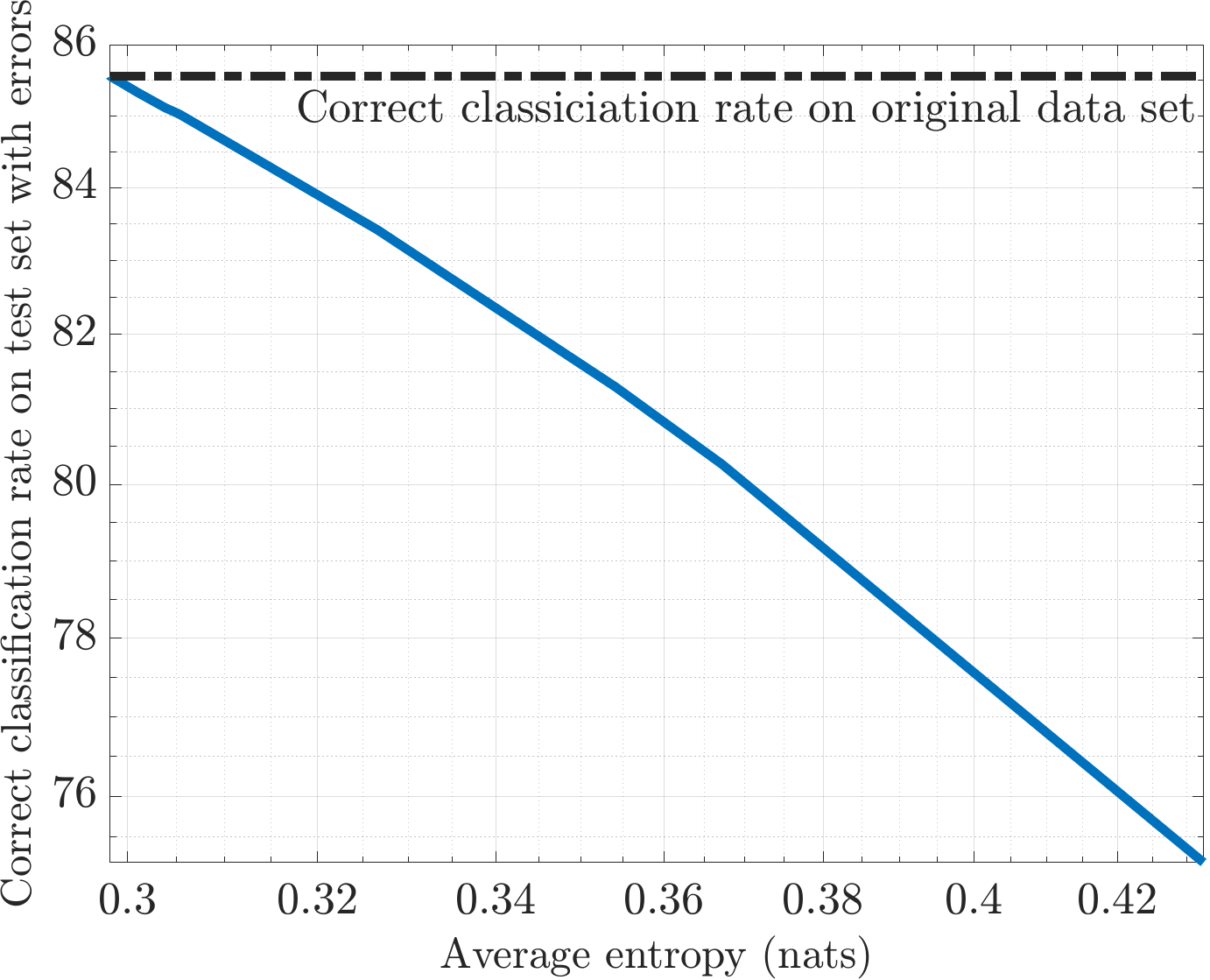}
        \label{fig:accuracy_vs_entropy}
    }

    \caption{Validation of assumptions $1$ and $2$}
    \label{fig:assumption_validation}

\end{figure}
Assumption $2$ is also not supported by a mathematical proof, and can fail in some specific cases (e.g., a very different data distribution in a test set). However, it is mild and valid in many operating conditions. Also, the cross-entropy is the typical loss for DNN training. In Fig. \ref{fig:assumption_validation}, we validate the two assumptions on the CIFAR-$10$ data set \cite{krizhevsky2009learning}, whose detailed description, along with the one of the trained CNN architecture, is provided in section \ref{sec:dataset_description} to lighten the reader. In particular, Fig. \ref{fig:entropy_vs_PER} shows the average entropy over the test set, as a function of the PER, while Fig. \ref{fig:accuracy_vs_entropy} shows the test accuracy as a function of the average entropy.
Finally, for the sake of complete definition of goal-effectiveness, we remind that the end-to-end inferencing delay is given by \eqref{delay_tot}.

Now, we can formalize the goal and its effectiveness for the investigated system, i.e. the edge inference service, coexisting with a legacy communication service. The goal is to obtain inference results within a maximum delay $D_{\max}$ from the time instant the classification request is issued\footnote{In this paper, with reference to a batch of patterns, we assume that the time instant of batch-based classification task generation coincides with the start of batch transmission in the uplink. This assumption can be easily generalized and will be investigated in future works.}, with goal value (cf. \eqref{goal_value_EI}) higher then a predefined threshold $\Theta_{\textrm{th}}$. Formally, the goal is achieved if the event $\{D_{\textrm{tot},t}\leq D_{\max}\}\cap \{\Theta_t\geq \Theta_{\textrm{th}}\}$ occurs, or, equivalently, a \textit{goal outage} is represented by one of the following  events: \textit{i)} $\{D_{\textrm{tot},t}> D_{\max}\}\cup \{\Theta_t\geq \Theta_{\textrm{th}}\}$,  \textit{ii) $\{D_{\textrm{tot},t}\leq D_{\max}\}\cup \{\Theta_t< \Theta_{\textrm{th}}\}$}, or \textit{iii)} $\{D_{\textrm{tot},t}> D_{\max}\}\cup \{\Theta_t< \Theta_{\textrm{th}}\}$. Then, we can define the goal-effectiveness as the probability of achieving the goal:
\begin{equation}\label{goal_effectiveness_EI}
    \mathcal{E}_{g}=\lim_{T\to\infty}\frac{1}{T}\sum_{t=0}^{T-1}\mathbb{E}\left\{\mathbf{1}_{\{\Theta_t\geq \Theta_{\textrm{th}}\}}\cdot\mathbf{1}_{\{D_{\textrm{tot},t}\leq D_{\max}\}}\right\},
\end{equation}

Now, let us note that the data rate in \eqref{data_rate} is a monotonic increasing function of the target PER $\gamma_t$, i.e., a higher PER tolerance allows $UE_g$ to transmit data at higher speed, which also means lower E2E delay (cf. \eqref{delay_comm}, \eqref{delay_tot}). This straightforward, yet, fundamental behaviour, and the goal-effectiveness definition in \eqref{goal_effectiveness_EI}, come with non-trivial consequences, among which we identify the following: the edge inference system can be either limited by the inference entropy, or by the E2E delay, depending on selected target PER and received interference by the DO system (i.e., wireless performance), as well as on edge processing delay (i.e., computation performance). Indeed, due to Assumption $1$, lower PER leads to better inference performance (from a classification confidence point of view), while it could be highly detrimental from an E2E delay, and, therefore, classification timeliness perspective (cf. \eqref{data_rate}). As a consequence, it is not necessarily more convenient to transmit classification input data with ultra high communication reliability (e.g., target PER below $10^{-7}$), but rather with the minimum level of communication reliability that \emph{jointly} guarantees the \textit{target level of classification confidence and its timeliness}. We argue that, based on the definition in \eqref{goal_effectiveness_EI}, a lower goal-effectiveness could be achieved as a result of ultra-reliable, yet, slower wireless communication, which is the purpose of the next section. 

The aim of the goal-oriented optimization approach is to find the right balance between the two measures, to attain the desired goal-effectiveness, entailing both inference timeliness and inference confidence (goal value). So far, we did not formalize the goal cost of the scenario under our investigation.

\subsection{The DO user data rate loss as goal cost}\label{sec:goal_cost_EI}
Let us recall that our aim is to assess performance in a scenario in which a GO system coexists with a DO system. In the considered scenario, the DO user, $UE_d$, uploads data (e.g., a video stream) with the aim of maximizing its data rate. However, the system is cooperative, and the DO user aims to maximize its data rate, without preventing the GO user from attaining the desired goal-effectiveness. Denoting by $\mathbf{w}_d\in\mathbb{C}^{M_d\times 1}$ the combining vector of $AP_d$, the received SINR at $AP_d$ at time $t$, reads as follows:
\begin{equation}
    \textrm{SINR}_{d,t}=\frac{|\mathbf{w}_{d,t}^H\mathbf{h}_{dd,t}|^2P_{d,t}}{N_0W+|\mathbf{w}_{d,t}^H\mathbf{h}_{dg,t}|^2P_{g,t}},
\end{equation}
and the data rate can be approximated by the Shannon formula, as we do not need to assess the PER performance for the DO system (however, generalizing this would be straightforward):
\begin{equation}
    R_{d,t}=W\log_2(1+\textrm{SINR}_{d,t}).
\end{equation}
Denoting by $\mathcal{R}_{d,\max}$ the maximum achievable average data rate by the DO system in the absence of the GO one and with transmission power $P_d = P_{d,\max}$, where, $P_{d,\max}$ stands for the maximum transmission power of $UE_d$, the goal cost can be defined as the \emph{average relative $UE_d$ data rate loss}, i.e., it can be written as follows (cf. \eqref{long_term}):
\begin{equation}\label{goal_cost_EI}
    \mathcal{C}=\frac{\mathcal{R}_{d,\max}-\mathcal{R}_{d}}{\mathcal{R}_{d,\max}}.
\end{equation}
The lower the DO user data rate is, the higher the goal cost is. In other words, the price to make the GO user achieve its goal with target goal-effectiveness, is paid by the DO user through its achievable uplink data rate. In the sequel, we assume that the DO user always has backlogged traffic, i.e., it continuously transmits and interferes with the GO system. This can be easily generalized.

Now, we have all elements describing goal-effectiveness and goal cost, in the investigated scenario. In the next section, to gain insights on relevant parameters to be controlled and optimized in the proposed system, we will first present a performance assessment, obtained through Monte Carlo simulations. Then, based on the acquired messages, a goal-oriented problem formulation, along with its solution, will be presented in section \ref{sec:problem_formulation}. We believe that this gentle introduction will help the reader understanding all ingredients of the system, and also the reasoning behind the goal-oriented approach, and the corresponding proposed optimization problem.
\subsection{Evaluation of the system without optimization}\label{sec:evaluation}
In this section, we evaluate the performance of the overall system, in which GO and DO systems coexist. To do so, Monte Carlo experiments are conducted, considering different realizations of wireless channels, computation delays, and pattern inference requests (i.e., number of patterns generated at each time slot). Results are obtained from $T=50000$ independent realizations of such parameters. The following assumptions hold for communication, computation, and inference settings.

\subsubsection{Wireless communications assumptions}
We consider two users (i.e., $UE_g$ and $UE_d$), two APs (i.e., $AP_g$ and $AP_d$), and an MEH, as in Fig. \ref{fig:CR_GO}. $UE_g$, $AP_g$, $UE_d$, and $AP_d$ are placed at $[5,0]$, $[5,20]$, $[8,0]$, $[8,20]$, respectively. Both UEs are equipped with a single antenna, and both APs employ a uniform linear array of $M_g=M_d=8$ elements. Maximal ratio combining is employed for the receive filters at both $AP_g$ and $AP_d$. Channel coefficients are generated as in \eqref{channel_matrix}, with path loss $\beta_{ij}$ with exponent $4$, while Rice fading with factor $K=3$ is considered. A different channel realization is extracted for each experiment from the described distribution. To evaluate the performance, we assume the transmit power of $UE_g$ to be fixed to $P_g=100$ mW, while we vary the transmit power of $UE_d$, choosing it in $[0,P_{c,\max}]$ W, with $500$ evenly spaced points, and $P_{c,\max}=200$ mW, in order to explore the performance as a function of $UE_d$ data rate (in each simulation, we will specify the selected value, whenever needed). The carrier frequency is $f_c=28$ GHz\footnote{https://www.bmwk.de/Redaktion/EN/Publikationen/Digitale-Welt/guidelines-for-5g-campus-networks-orientation-for-small-and-medium-sized-businesses.pdf}, with total bandwidth $W=1$ GHz, fully shared among $UE_g$ and $UE_d$, while the noise power spectral density fixed to $N_0=-174$ dBm/Hz, with a $3$ dB noise figure at the receiver. The transmission of $UE_g$ is organized in packets, each of equal size of $32$ Bytes. The PER is chosen from the vector $\mathbf{\gamma}=[10^{-7},10^{-6},10^{-5},10^{-4},2\times 10^{-4},4\times 10^{-4},8\times 10^{-4},10^{-3},2\times 10^{-3},4\times 10^{-3},8\times 10^{-3},10^{-2},2\times 10^{-2}]$, to explore performance in terms of communication reliability. Therefore, at time $t$, given a PER $\gamma_t$, the transmitted packets are considered to be incorrectly received with probability $\gamma_t$ chosen from the vector (and specified for each simulation, whenever needed). When a packet is not correctly received, no retransmission is requested, and the bits within the packet payload are randomly chosen. 
\subsubsection{Data set and inference assumptions}\label{sec:dataset_description}
We evaluate the performance on the CIFAR-$10$ data set \cite{krizhevsky2009learning}, which consists of $32\times32$ pixel RGB images, with $50000$ patterns in the training set, and $10000$ patterns in the test set. We resize each image to have $64\times 64$ pixel RGB images, representing each of the three base colors with $32$ bits. We assume that a pre-trained CNN is pre-onboarded at the MEH and ready to provide classification results. We consider batches $\mathcal{B}_t$ of equal size of $N_t=20$ patterns (cf. \eqref{delay_comm}, \eqref{batch_entropy}) for each realization. Therefore, considering the packet size of $32$ Bytes and the image size, the total number of packets transmitted for each experiment (i.e., time slot) is $\frac{64^2\times3\times 32\times 20}{256}=30720$. We assume that a state of the art model\footnote{https://appliedmachinelearning.wordpress.com/2018/03/24/achieving-90-accuracy-in-object-recognition-task-on-cifar-10-dataset-with-keras-convolutional-neural-networks/} is pre-trained and pre-uploaded in the MEH. The model is trained using Matlab$^{\text{\textregistered}}$. For simplicity, we also assume a fixed blocklength $n_g=64$ Bytes.
\subsubsection{Computation delay assumption}\label{sec:comp_delay_evaluation}
\begin{figure}[t]
    \centering
    \includegraphics[width =\columnwidth]{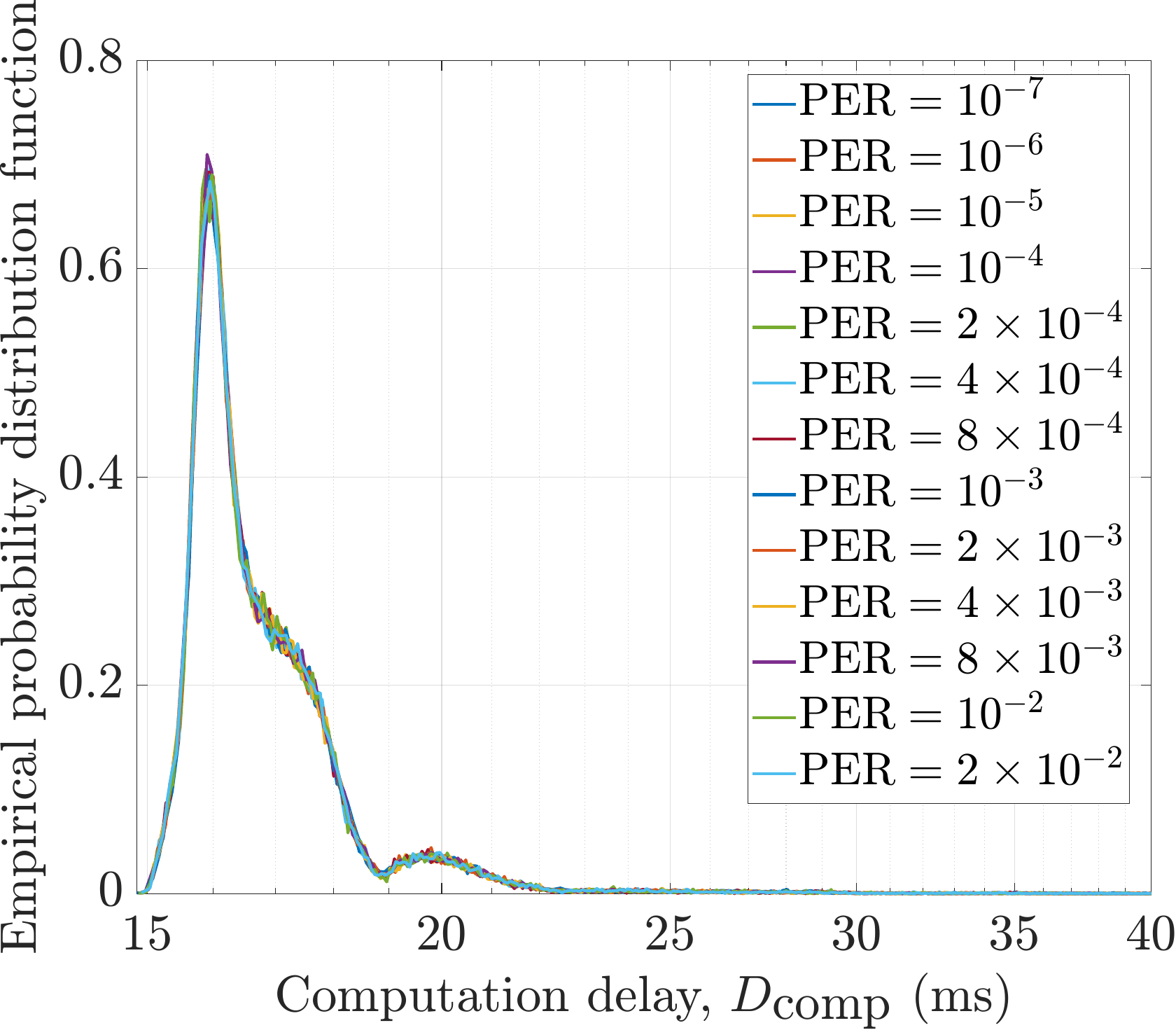}
    \caption{Empirical probability density function of computing delay, estimated through real experiments}
    \label{fig:delay_pdf}
\end{figure}
In the absence of a model for the computing delay, we have empirically built a delay distribution from real-world experiments. The inference runs on an GPU NVIDIA$^{\text{\textregistered}}$ Tesla$^{\text{\textregistered}}$ V100. The CPU characteristics are the following: Intel$^{\text{\textregistered}}$ Xeon$^{\text{\textregistered}}$ Gold 6244 CPU@3.6 GHz, with $4$ cores and $8$ GB of memory. Then, we run batch inference for $T=50000$ independent realizations, we save the computing time, and we retrieve a computing delay distribution, from which we can extract a random realization when evaluating performance. For the interest of the reader, the estimated probability density function of the computing delay is shown in Fig. \ref{fig:delay_pdf}, for all values of PER, as (a priori) the distribution may change as a function of the PER. However, from the figure, it can be noted that the distribution is stable over PER values, showing only slight variations, probably caused by fluctuations in the CPU usage and background processes. Therefore, in the sequel, we will consider the computation delay distribution as independent from the target PER.

We will now present the results in terms of goal-effectiveness, as per its definition in \eqref{goal_effectiveness_EI}. Let us recall that the goal-effectiveness generally entails the goal value and other system constraints. For instance, in the case of edge inference, the goal value is represented by the NREI (cf. \eqref{goal_value_EI}), and the other system constraint is represented by the E2E delay (cf. \eqref{delay_tot}, \eqref{delay_comm}, Fi. \ref{fig:delay_pdf}). However, for the sake of smooth exposition, we will first present two separate results, i.e., the goal-effectiveness by only taking into account the goal value - NREI (c.f. \eqref{goal_value_EI}), and the goal-effectiveness by only taking into account the delay, respectively. We believe that this gentle introduction will help us commenting the results, and the reader comprehending the conveyed message, via a clear explanation of the different sources of goal outages. The two notions of goal-effectiveness will be then merged in Section \ref{sec:goal_effect_merged} as a final result, to coherently consider the complete definition in \eqref{goal_effectiveness_EI}. Afterwards, an optimization problem will be proposed in section \ref{sec:problem_formulation}.
\subsubsection{The goal-effectiveness from the goal value perspective}
\begin{figure}[htb!]
    \centering
    \includegraphics[width =\columnwidth]{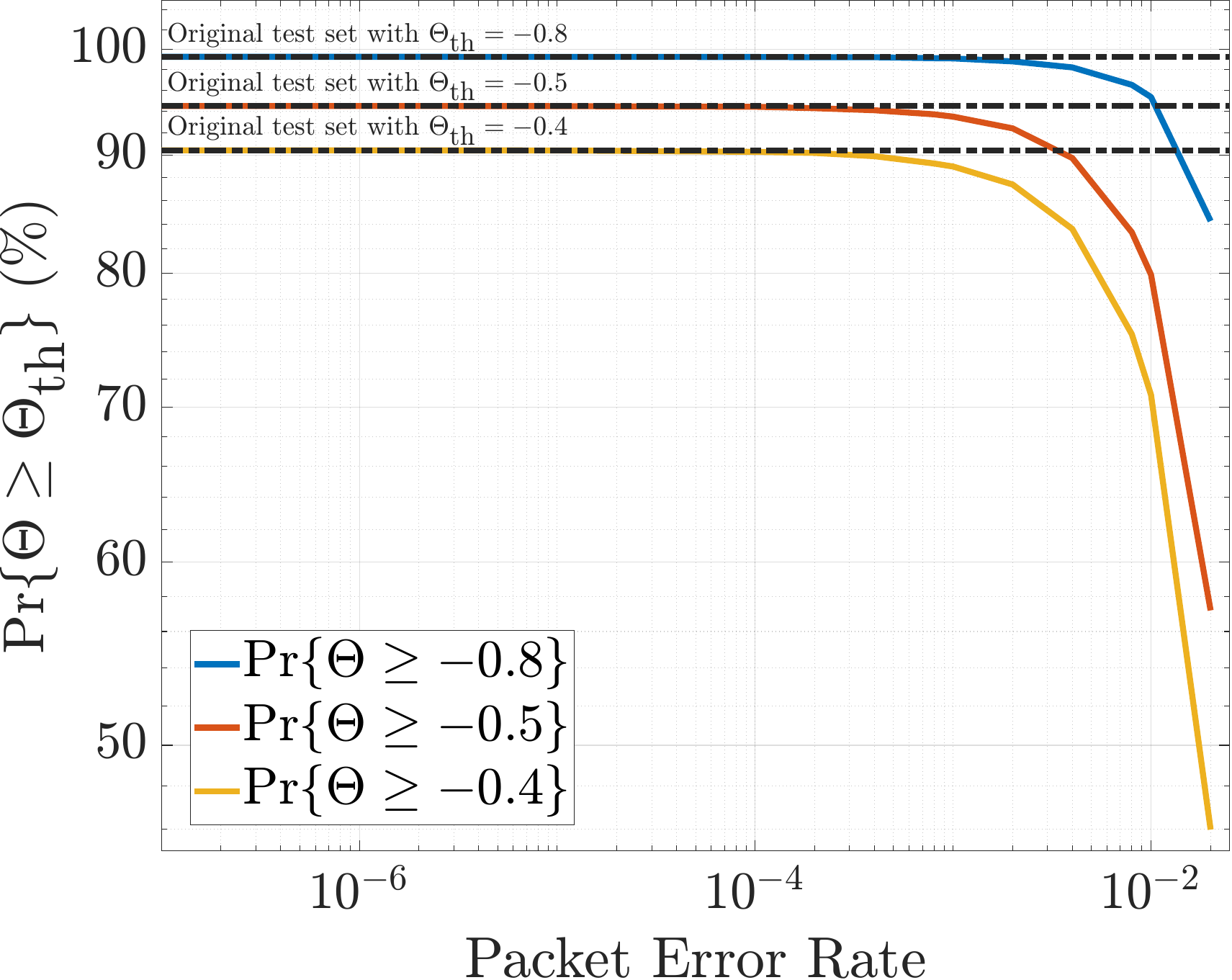}
    \caption{Probability of goal value (NREI cf. \eqref{goal_value_EI}) being above a predefined threshold, as a function of PER, for different threshold values}
    \label{fig:entropy_effec_vs_PER}
\end{figure}
Let us evaluate the performance of goal-effectiveness, only from the point of view of the goal value, i.e., the first term on the right hand side of \eqref{goal_effectiveness_EI} (the inference reliability - negative relative entropy increase- NREI). In other words, we consider a received batch result with goal value higher than the predefined threshold, as a goal achievement, even if results are issued after the deadline $D_{\max}$. Let us first notice that the goal value, as per its definition in \eqref{goal_value_EI}, is only affected by the PER, i.e., the errors generated by the wireless communication between the GO user and the GO AP. Then, in Fig. \ref{fig:entropy_effec_vs_PER}, we show the goal-effectiveness related to the goal value as a function of the PER, for different thresholds $\Theta_{\textrm{th}}$. First, we can notice how the probability of the goal value being above a threshold, decreases as the PER increases, as more errors occur throughout the uplink communication phase. This result is also in line with Assumption $1$ (Fig. \ref{fig:entropy_vs_PER}). Also, by increasing the threshold $\Theta_{\textrm{th}}$ (recall that the average NREI assumes non positive values), performance degrades, as expected. 

As already mentioned, this result does not take into account the delay, which is a fundamental part of the goal (i.e., the edge inference service). In the following, we consider the goal-effectiveness from the delay perspective, to then merge the two perspectives in Section \ref{sec:goal_effect_merged}.
\subsubsection{The goal-effectiveness from the delay perspective}
\begin{figure}[htb!]
    \centering
    \includegraphics[width =\columnwidth]{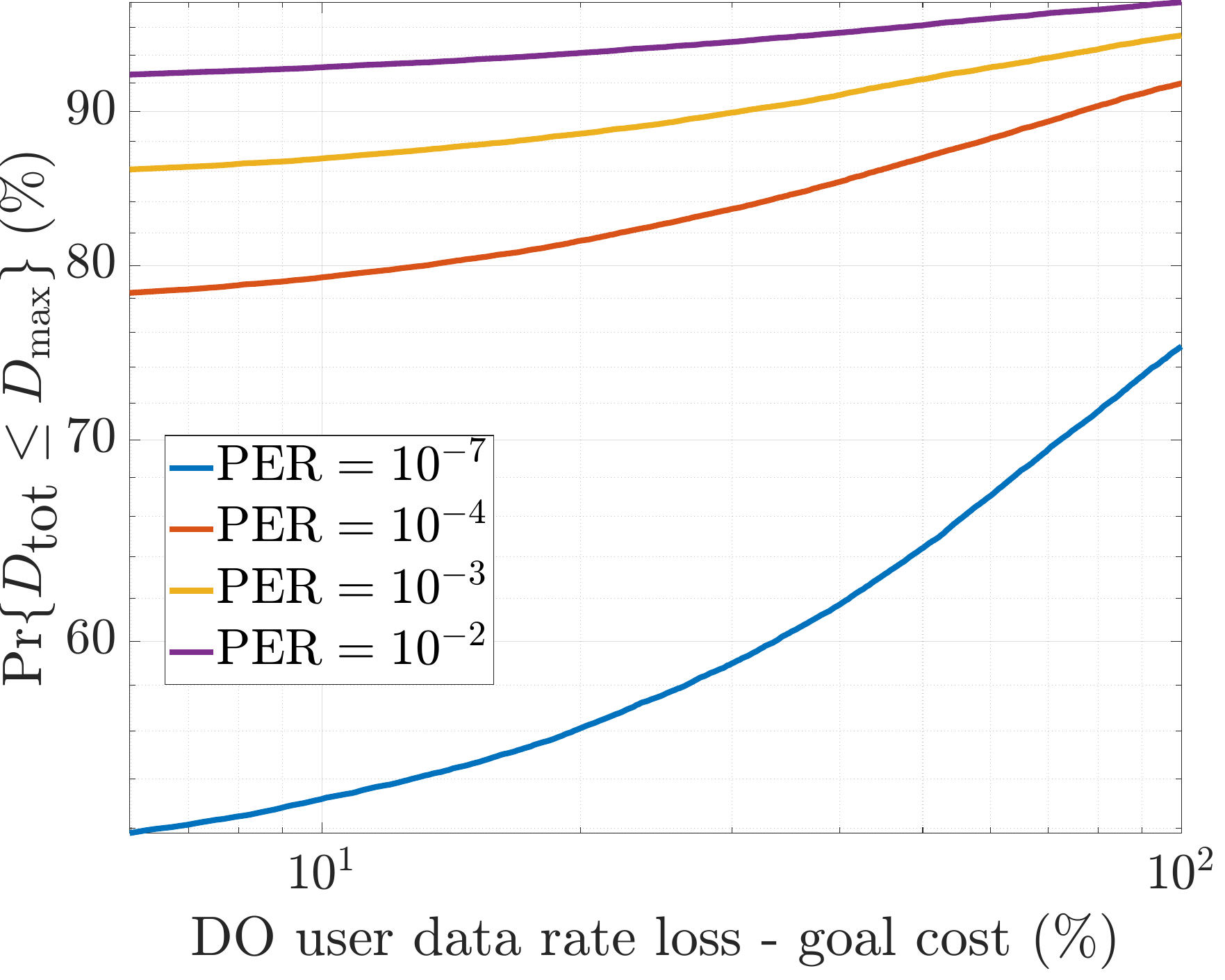}
    \caption{Probability of guaranteeing the delay constraint vs. goal cost (DO user relative data rate loss)}
    \label{fig:delay_effec_vs_rate}
\end{figure}
\begin{figure}[htb!]
    \centering
    \includegraphics[width =\columnwidth]{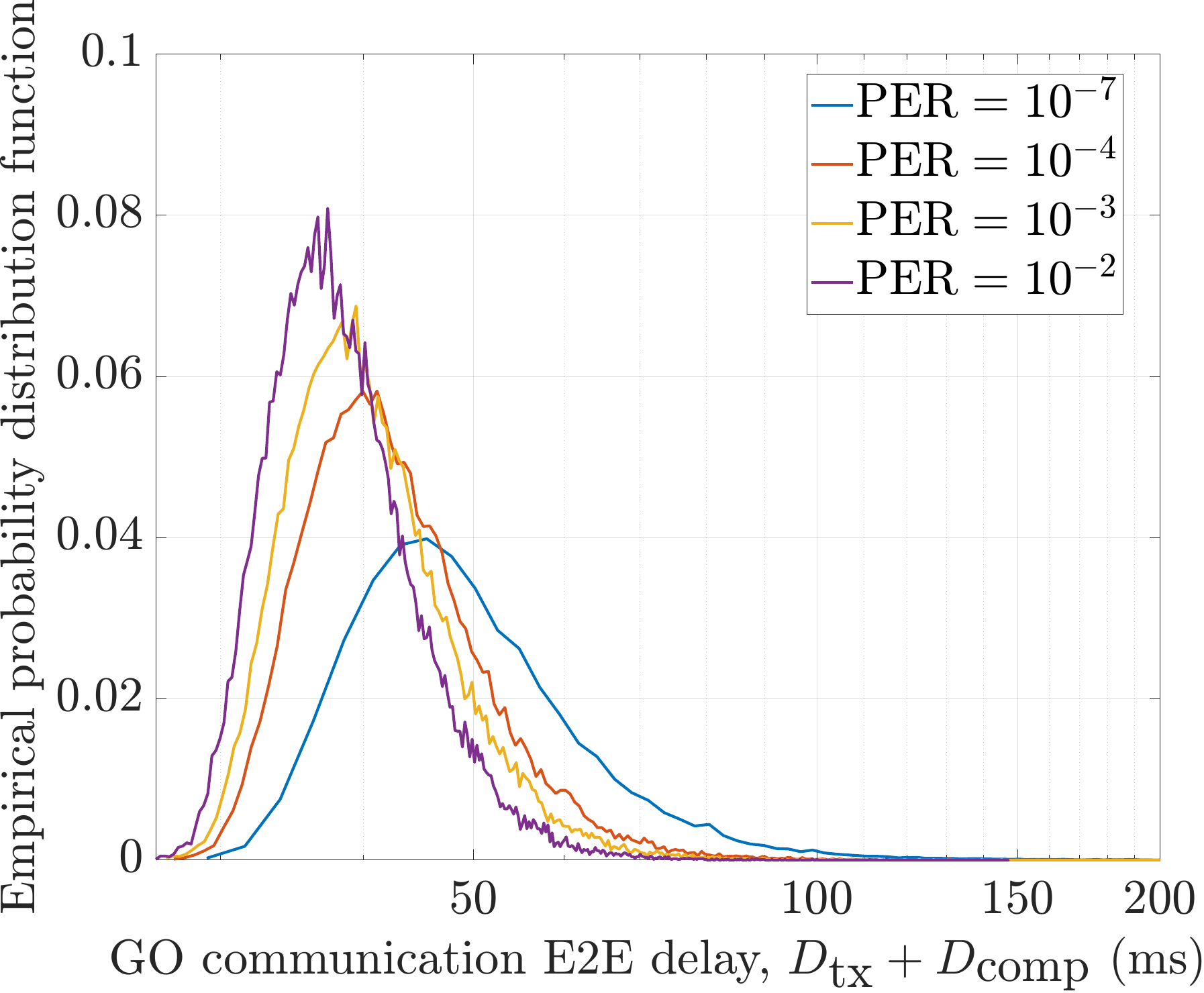}
    \caption{Empirical probability density function of E2E delay for goal cost $\mathcal{C}\approx6$\% (left hand points of Fig. \ref{fig:delay_effec_vs_rate})}
   \label{fig:tot_delay_pdf}
\end{figure}
First, it is worth noting that the E2E delay depends on the computation delay, the PER and the interference (cf. \eqref{data_rate}, \eqref{delay_comm}). In Section \ref{sec:comp_delay_evaluation} (Fig. \ref{fig:delay_pdf}), we empirically illustrated the independence between PER and computation delay. On the other hand, the communication delay is highly dependent on PER and received interference. Indeed, at fixed channel conditions and GO user transmit power $P_g$, by increasing the DO user's transmit power $P_d$, the data rate of the latter increases, thus, generating more interference to the GO system, which causes higher delay, in case of fixed target PER. To show the effect of the E2E delay on the effectiveness, we consider a delay threshold $D_{\max}=50$ ms, and we plot, in Fig. \ref{fig:delay_effec_vs_rate}, the probability for a batch to be classified by the deadline, as a function of the DO user data rate loss (i.e., the goal cost in \eqref{goal_cost_EI}), for a subset of target PER $[10^{-7},10^{-4},10^{-3}, 10^{-2}]$ (for ease of readability). This is obtained by increasing the DO user transmit power $P_d$ in $[0,200]$ mW. As we can notice, contrarily as before, lower PER leads to lower effectiveness (as the effectiveness only entails delay, i.e., the GO user does not care about receiving high entropy results, provided that they are received within the predefined E2E delay). This result is due to the strong dependence of the communication delay (and thus the E2E delay) on PER and interference. Indeed, as an example, let us show, in Fig. \ref{fig:tot_delay_pdf}, the empirical probability density function of the GO system E2E delay (communication and computation), for the left hand point of Fig. \ref{fig:delay_effec_vs_rate}, i.e., for a single value of $P_{c}=200$ mW. As we can notice, the distribution of the delay experiences longer tails for lower PER, as expected from \eqref{data_rate}. From the last results, we can easily conclude that:
\begin{itemize}
    \item From a goal value perspective, the goal-effectiveness is only affected by the PER (although not strongly, depending on the PER value), and a higher PER (i.e., lower communication reliability) leads to lower (partial) goal-effectiveness; also, from $\gamma=10^{-7}$ to $\gamma=10^{-3}$, stable performance is experienced.
    \item From an E2E delay perspective, the goal-effectiveness is affected by the PER and the DO user interference, and higher PER (i.e., lower communication reliability) leads to higher (partial) goal-effectiveness.
\end{itemize}
The goal of this work is to analyze and optimize performance, by taking into account the overall definition of a goal (and goal-effectiveness), entailing the goal value and the end-to-end delay, but also the goal cost. Therefore, in what follows, we finally present the results in terms of goal-effectiveness, as per its overall definition in \eqref{goal_effectiveness_EI}, together with the corresponding goal cost.
\subsubsection{The goal-effectiveness and its dependence on PER, interference, and goal cost}\label{sec:goal_effect_merged}
\begin{figure*}[htb!]
    \centering
    \subfloat[$\Theta_{\textrm{th}}=-0.8$]{
        \includegraphics[width=0.343\textwidth]{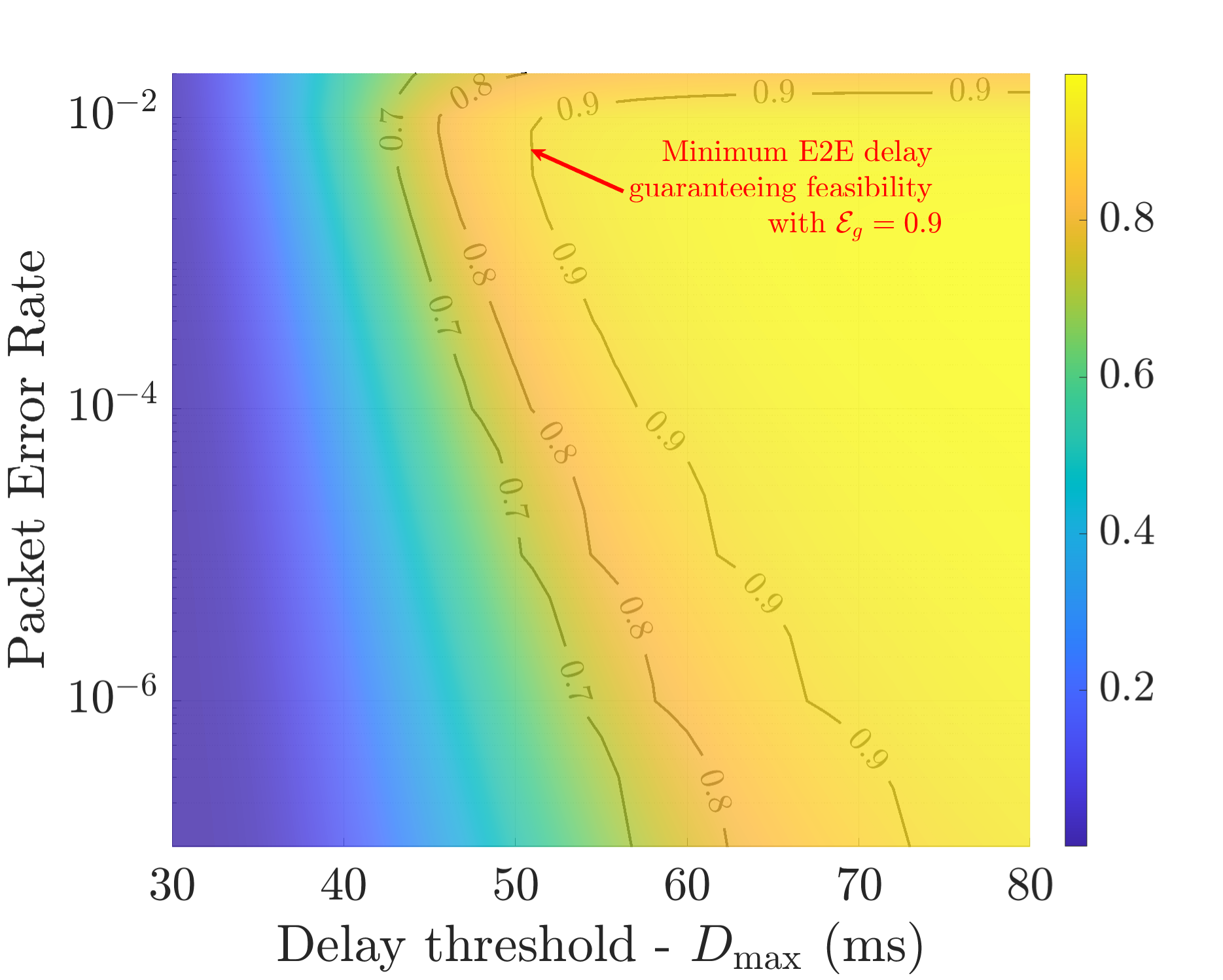}
        \label{fig:heatmap_entropy_80}
    }
    \subfloat[$\Theta_{\textrm{th}}=-0.5$]{
        \includegraphics[width=0.335\textwidth]{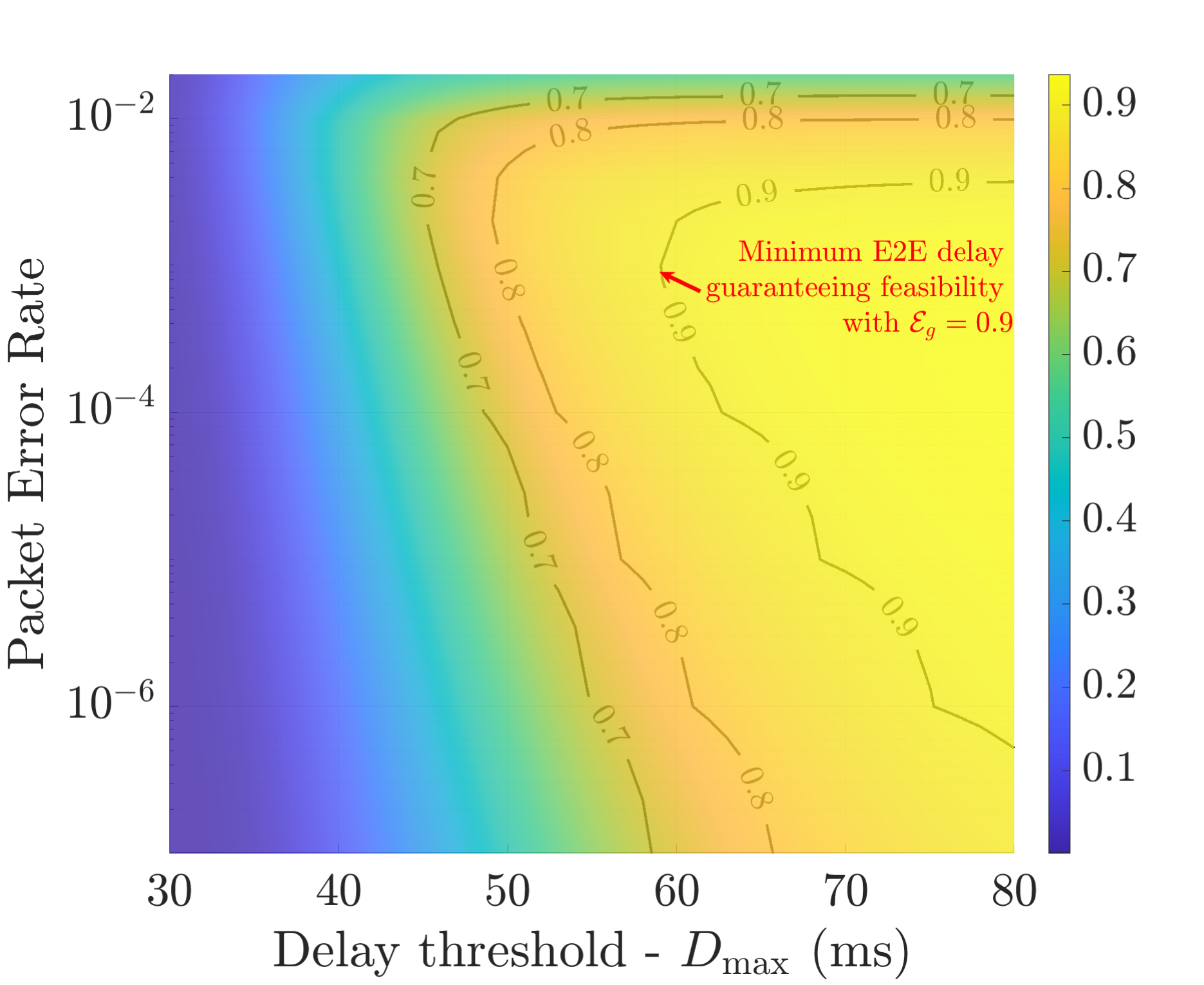}
        \label{fig:heatmap_entropy_50}
    }
    \subfloat[$\Theta_{\textrm{th}}=-0.4$]{
        \includegraphics[width=0.343\textwidth]{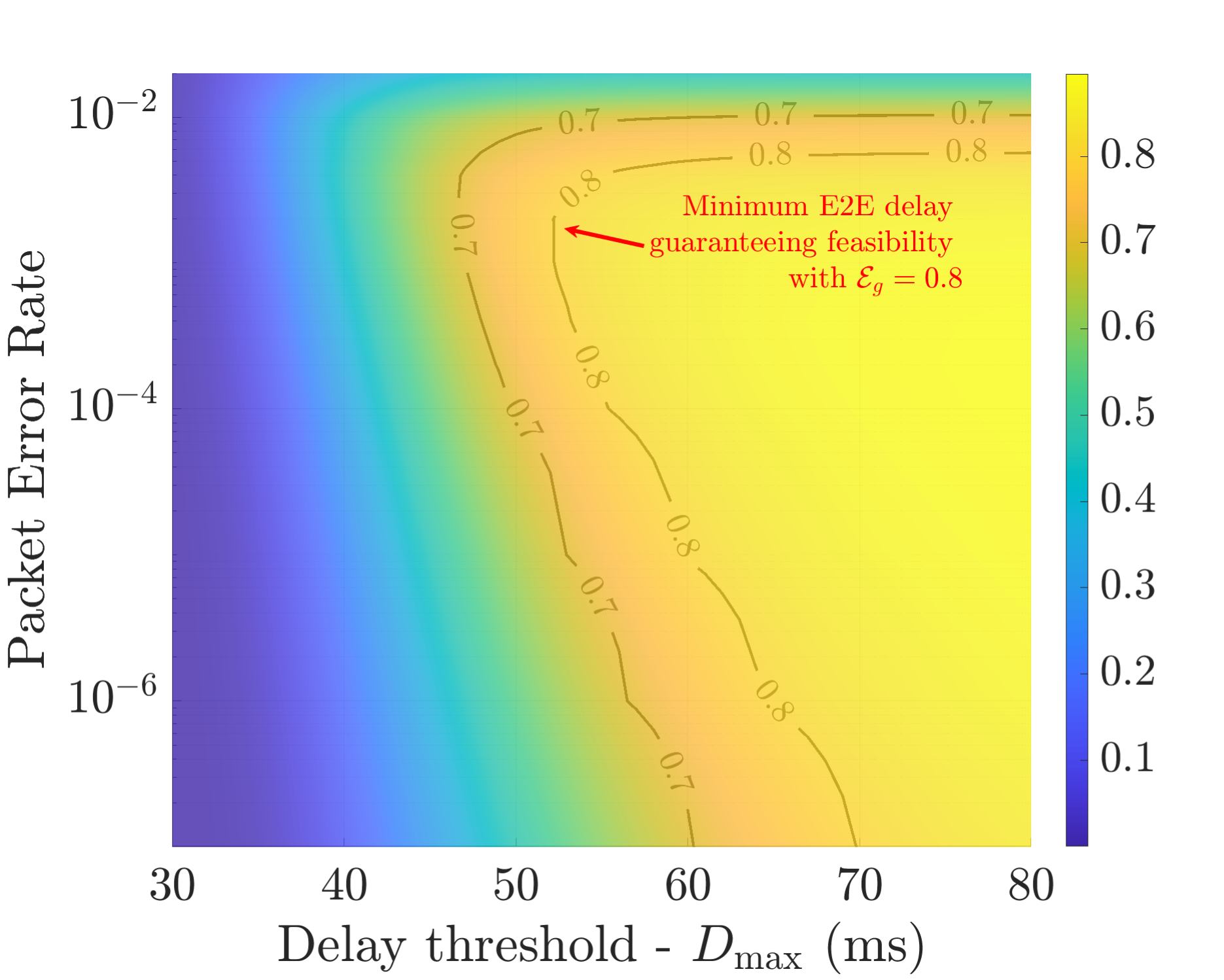}
        \label{fig:heatmap_entropy_40}
    }
    
        \subfloat[$\Theta_{\textrm{th}}=-0.8$]{
        \includegraphics[width=0.34\textwidth]{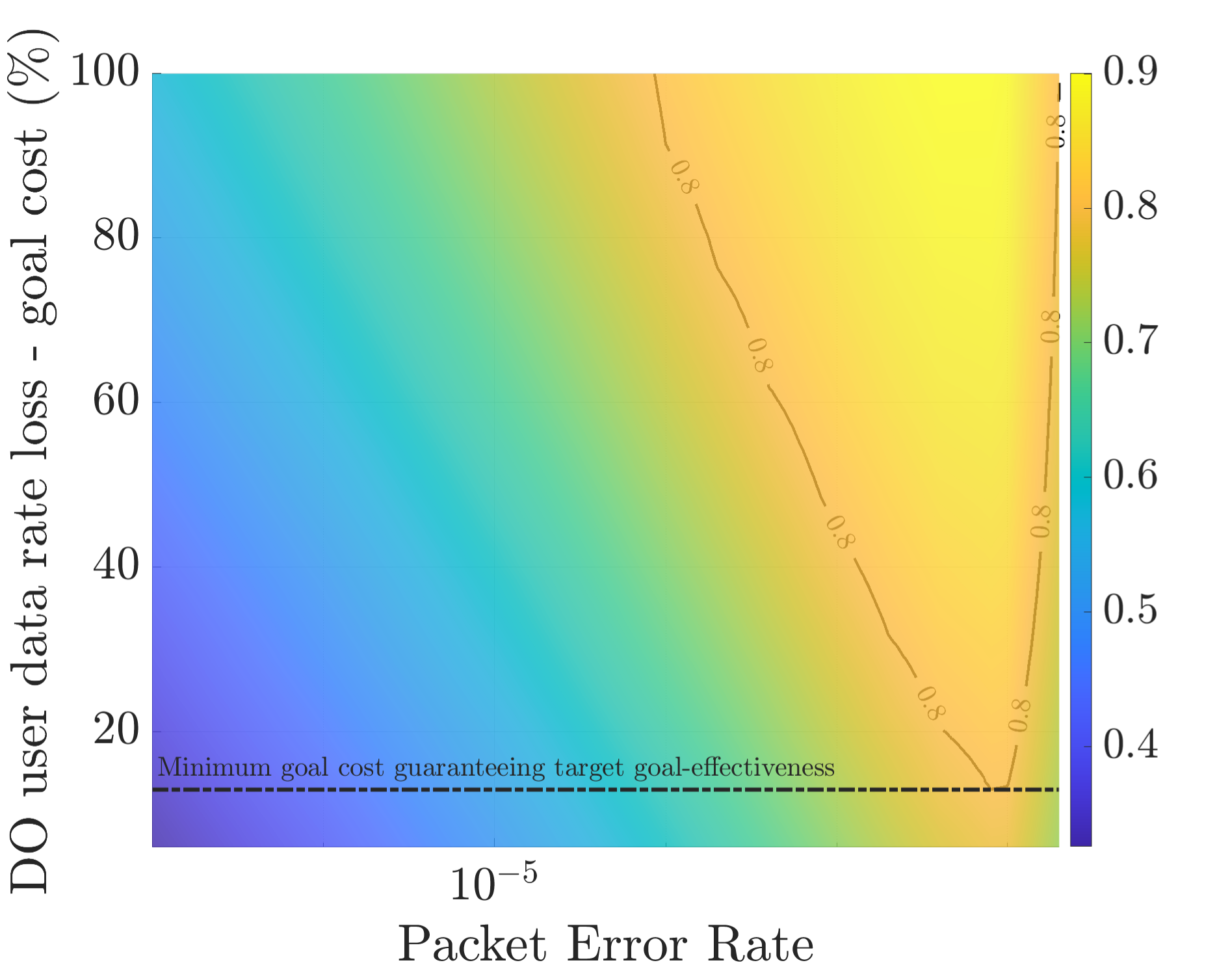}
        \label{fig:cost_PER_effectiveness_80}
    }
    \subfloat[$\Theta_{\textrm{th}}=-0.5$]{
        \includegraphics[width=0.34\textwidth]{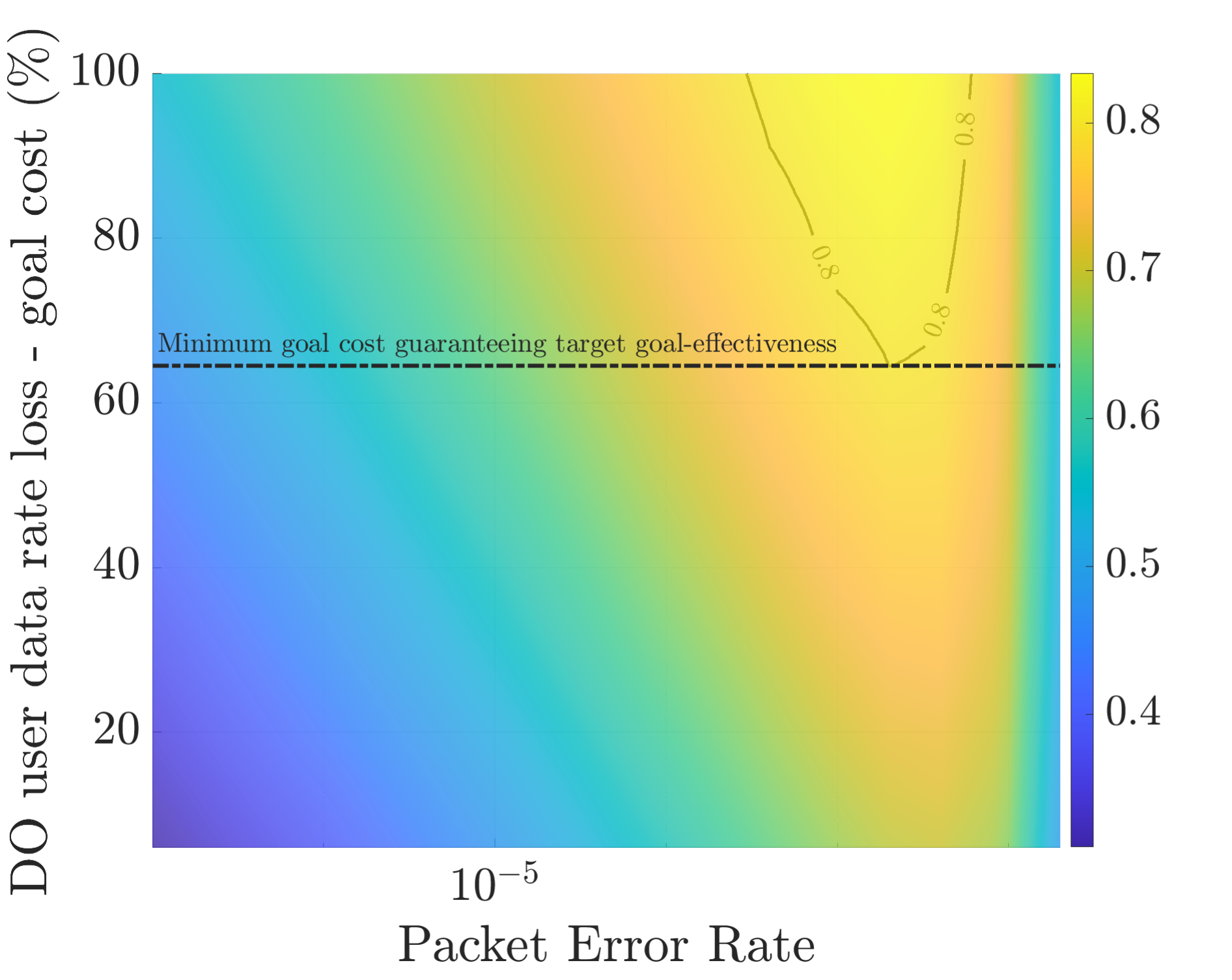}
        \label{fig:cost_PER_effectiveness_50}
    }
    \subfloat[$\Theta_{\textrm{th}}=-0.4$]{
        \includegraphics[width=0.34\textwidth]{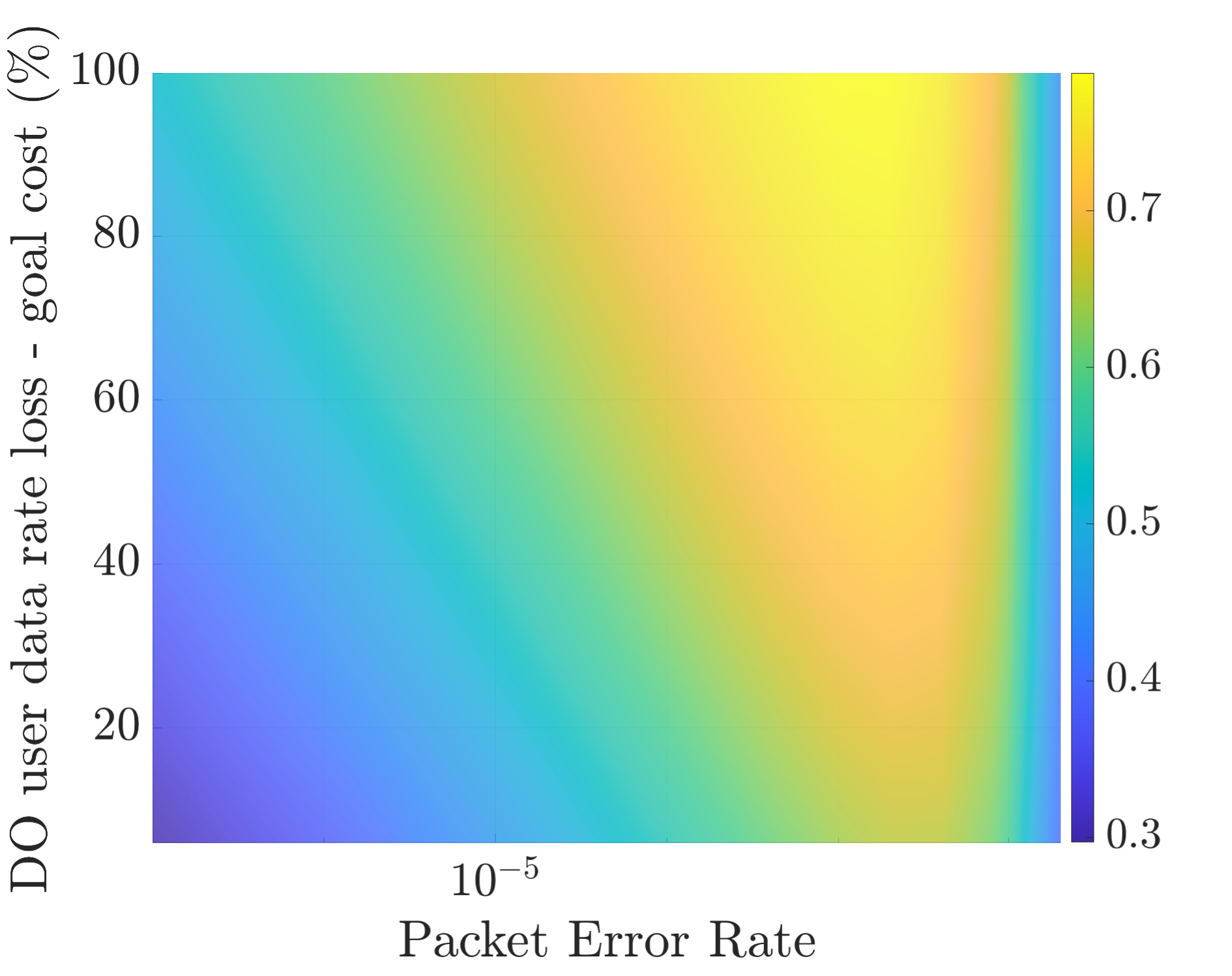}
        \label{fig:cost_PER_effectiveness_40}
    }
    \caption{(a,b,c) Goal-effectiveness heat maps as a function of GO communication PER and E2E delay constraint, for fixed goal cost, and different goal value thresholds $\Theta_{\textrm{th}}$; (d,e,f) Goal-effectiveness heat maps as a function of goal cost (relative DO user data rate loss) and target PER of GO system, for fixed GO communication E2E delay threshold ($45$ ms) and different goal value thresholds $\Theta_{\textrm{th}}$}
    \label{fig:heatmaps}
\end{figure*}

The goal-effectiveness in \eqref{goal_effectiveness_EI} depends on two performance indicators: \textit{i)} the goal value (represented by the NREI), which in our setting directly and exclusively depends on the PER, and \textit{ii)} the E2E delay, which depends on the PER, the interference caused by the DO user, and the remote computation delay. As a first joint result, in Fig. \ref{fig:heatmaps}, we show heat maps representing the goal-effectiveness (i.e., the goal-effectiveness is represented by the different colors from dark blue to yellow) as a function of PER ($y$-axis) and maximum delay threshold $D_{\max}$ ($x$-axis), for a fixed DO user data rate (i.e., fixed $P_d=200$ mW), and for different goal value thresholds in three different plots, i.e. $\Theta_{\textrm{th},1}=-0.8$, $\Theta_{\textrm{th},1}=-0.5$, and $\Theta_{\textrm{th},1}=-0.4$, shown in Figs. \ref{fig:heatmap_entropy_80}, 
\ref{fig:heatmap_entropy_50}, and \ref{fig:heatmap_entropy_40}, respectively (we remind that a higher threshold indicates a stricter constraint in the sense of goal achievability). Note that the heat maps are interpolated a posteriori for better visualization, while results have been obtained for the sub-selected PER and $D_{\max}$ values. In these figures, we also show different goal-effectiveness thresholds (i.e. $\mathcal{E}_{g,\textrm{th}}=0.7$, $\mathcal{E}_{g,\textrm{th}}=0.8$, $\mathcal{E}_{g,\textrm{th}}=0.9$), through contours plots. All the points interior to these contour plots represent a \textit{goal feasibility region} for each respective goal-effectiveness constraint, i.e., all the combinations of (fixed) PER and end-to-end delay, whose corresponding goal-effectiveness exceeds a predefined threshold. However, these regions are subject to the fact that there is no adaptation of transmit power and target PER across time; whereas, as we will show in the sequel, much better results can be obtained in the optimized setting. From these plots, we can make the following considerations:
\begin{itemize}
        \item The goal-effectiveness increases as the delay threshold (y-axis) increases (for each fixed target PER), while it does not necessarily decrease as a function of the PER, as expected and shown before in the disjoint plots, creating \textit{goal-effectiveness feasibility regions}, whose surface depends on $\mathcal{E}_{g,\textrm{th}}$.
    \item The goal-effectiveness feasibility regions shrink as the goal value (NREI) threshold increases (see the difference between surface extensions in Figs. \ref{fig:heatmap_entropy_80}, \ref{fig:heatmap_entropy_50}, \ref{fig:heatmap_entropy_40} - looking at \ref{fig:heatmap_entropy_40}, we can even notice that no combination of fixed PER and $D_{\max}$ guarantees goal-effectiveness above $0.9$ for $\Theta_{\textrm{th}}=-0.4$ in this setting).
    \item Given a goal-effectiveness requirement, there always exists a minimum E2E delay threshold guaranteeing feasibility; whereas, below this threshold, it is infeasible (for any PER) to guarantee the requirement (examples of this point are shown by the red arrows in the figures). Moreover, each target PER experiences a different minimum $D_{\max}$ that can be guaranteed. The lower is the PER, the higher the minimum feasible delay is.
    \item As the delay threshold decreases, the feasible region in terms of PER also shrinks, i.e., with a lower delay constraint, higher PER values are needed to guarantee effectiveness; however, this is not always feasible due to the goal value constraint (see, e.g., Fig. \ref{fig:heatmap_entropy_40}).
    \item As a consequence of the previous remark, as the PER decreases ($y$-axis), the minimum delay threshold to guarantee a target goal-effectiveness increases, i.e., to guarantee lower PER (reliable communication), $UE_g$ needs more time for offloading, resulting in more frequent delay outages. 
\end{itemize}
Let us now focus on Figs. \ref{fig:cost_PER_effectiveness_80}, \ref{fig:cost_PER_effectiveness_50}, \ref{fig:cost_PER_effectiveness_40}, in which we plot the goal-effectiveness, as a function of DO user data rate loss in \eqref{goal_cost_EI}, i.e., the goal cost ($y$-axis) and the PER ($x$-axis), for a fixed $D_{\max}=45$ ms, and different goal value (NREI) thresholds across figures, namely the same as the previous results. Also, a contour plot representing the $80$\% goal-effectiveness feasibility region is shown in all figures. These plots are obtained by also varying the DO user transmit power (within the range $[0,200]$ mW). From these results, we can appreciate the goal feasibility region, also as a function of the goal cost (i.e., the aim of this paper), and we can draw the following conclusions:
\begin{itemize}
    \item Again, the goal-effectiveness feasibility region is a surface, i.e. there are multiple solutions guaranteeing the goal-effectiveness constraint.
    \item While the above consideration holds, there exists a minimum goal cost solution, i.e., the minimum cost needed to achieve the target goal-effectiveness. The latter is the lowest point of the contour plots representing the effectiveness thresholds, and is represented by the black horizontal dashed lines in each plot.
    \item By increasing the goal value threshold (i.e., across different figures), the feasibility region shrinks as before and, as an additional observable effect, the minimum goal cost increases (e.g., above $60$\% of DO user data rate loss in Fig. \ref{fig:cost_PER_effectiveness_50}). In other words, the stricter the constraint in terms of goal value is, the higher is the minimum goal cost able to guarantee effectiveness.
    \item In certain conditions (see Fig. \ref{fig:cost_PER_effectiveness_40}) desired values of goal-effectiveness are not attainable (e.g., goal-effectiveness above $80$\%)
\end{itemize}
Interestingly, these results relate the performance of a legacy (DO) and a GO communication system interfering with each other, showing non trivial outcomes, which suggest that higher communication reliability does not necessarily imply higher goal-effectiveness. Therefore, they represent the basis to formulate a goal-oriented optimization problem in the next section. Indeed, the aim of a goal-oriented resource orchestration framework is to move within the goal-effectiveness feasibility region, possibly finding the lowest cost in such region. 

Then, from these results and their corresponding conclusions, it is straightforward to formulate a goal-oriented resource allocation problem, involving the variables that mostly affect goal-effectiveness and goal cost: \textit{i)} the PER of GO communication, and \textit{ii)} the DO user transmit power, in this work. Other variables can be taken into account, which represents further research directions on this topic.

\section{Problem formulation \& solution}\label{sec:problem_formulation}

The aim of this section is to propose a resource allocation policy able to minimize the  $UE_d$ data rate loss (cf. \eqref{goal_cost_EI}), subject to a goal-effectiveness constraint of the GO system. Of course, minimizing \eqref{goal_cost_EI} is equivalent to maximizing the average data rate of $UE_d$. Then, following the general formulation in \eqref{general_prob_form}, the edge inference problem can be formulated as follows:
\begin{align}\label{problem_formulation_EI}
    &\hspace{0 cm}\underset{\{\mathbf{\Psi}_t\}_t}{\max}\quad\mathcal{R}_d:=\lim_{T\to\infty}\frac{1}{T}\sum_{t=0}^{T-1}\mathbb{E}\{R_{d,t}\}\\
    &\text{subject to}\nonumber\\
    &(a)\;\lim_{T\to\infty}\frac{1}{T}\sum_{t=0}^{T-1}\mathbb{E}\left\{\mathbf{1}_{\{\Theta_t\geq \Theta_{\textrm{th}}\}}\cdot\mathbf{1}_{\{D_{\textrm{tot},t}\leq D_{\max}\}}\right\}\geq \mathcal{E}_{g,\textrm{th}}\nonumber,\\
    &(b)\; P_{d,t}\in\mathcal{P}_d,\;\forall t\nonumber,\quad (c)\; \gamma_t\in \Gamma_g,\;\forall t,\nonumber
\end{align}
where $\mathbf{\Psi}_t=[P_{d,t},\gamma_t]$ is the action set, i.e., the optimization variables, involving DO user transmit power and target GO user PER. The constraints of the problem have the following meaning: $(a)$ the goal-effectiveness of the GO system is higher than a predefined threshold; $(b)$ the instantaneous transmit power of $UE_d$ is chosen from a discrete set $\mathcal{P}_d$ involving a minimum value of transmission power ($0$ in this case) and a maximum value being equal to $P_{d,\max}$; $(c)$ the target PER of the GO system is chosen from a discrete set $\Gamma_g$. Also, we make the following assumptions:
\begin{enumerate}
    \item The goal is achievable, i.e., problem \eqref{problem_formulation_EI} is feasible.
    \item The optimization is performed at the MEH, which is provided with the needed connect-compute instantaneous information, as specified here below.
    \item All effective channels (i.e., including the receive filters) are perfectly known instantaneously, while their statistics are unknown in advance.
    \item The computation delay at the current time slot is estimated and known with high confidence, i.e., we assume the computation delay is known at time $t$. 
    \item The GO user has no buffered data, but it is able to accept all data patterns generated at time $t$, to be transmitted to the MEH.
    \item The DO user always has backlogged traffic, i.e., it continuously transmits and interferes with the GO system
    \item All thresholds (delay, entropy, effectiveness) are known in advance, i.e., they are requested as part of a service level agreement.
\end{enumerate}

Problem \eqref{problem_formulation_EI} is difficult to solve, due to the lack of distribution knowledge of wireless channels (i.e., only instantaneous realizations are observed) and computation delays at the MEH. Therefore, we hinge on the tools of Lyapunov stochastic network optimization to solve the problem on a per-slot basis, with instantaneous observations of context parameters. To this end, let us first introduce the concept of \textit{virtual queue} \cite{Neely10}, whose aim is to keep track of constraint violations, and take specific actions to drive the system towards desired operating modes. In particular, given a long-term constraint written as 
\begin{equation}\label{long_term_cons}
    \mathcal{X}:=\lim_{T\to\infty}\frac{1}{T}\sum_{t=0}^{T-1}\mathbb{E}\left\{x_t\right\}\leq x_{\textrm{th}},
\end{equation}
we can define an associated virtual queue that evolves as follows over successive time slots:
\begin{equation}\label{generic_virtual_queue}
    Z_{t+1}=\max(0,Z_t+x_t-x_{\textrm{th}}),
\end{equation}
The evolution of this mathematical model is straightforward: the size of the virtual queue grows whenever the constraint is violated, and it decreases otherwise. Interestingly, as easily proved in \cite[Section 4.4]{Neely10} constraint \eqref{long_term_cons} is guaranteed, if the associated virtual queue is \textit{mean-rate stable}, i.e., 
\begin{equation}
    \lim_{T\to\infty}\frac{1}{T}\mathbb{E}\left\{Z_T\right\}=0.
\end{equation}
To ensure the mean rate stability of the virtual queue, it is sufficient to guarantee that the so-called \textit{Conditional Lyapunov Drift (CLD)} is bounded by a finite constant at each slot. Therefore, let us introduce the CLD, by first defining the Lyapunov function \cite{Neely10}
\begin{equation}\label{LF}
    L(Z_t)=\frac{1}{2}Z_t^2,
\end{equation}
which is a measure of the congestion state of the system in terms of the defined virtual queue(s). From \eqref{LF}, the CLD is defined as follows:
\begin{equation}
    \Delta_t(Z_t)=\mathbb{E}\left\{L(Z_{t+1})-L(Z_t)|Z_t\right\},
\end{equation}
i.e., it is the expected variation of the Lyapunov function over two successive time slots. Guaranteeing that $\Delta_t(Z_t)\leq B$, with $B$ a finite constant, also ensures the mean-rate stability of $Z_t$, and therefore constraint \eqref{long_term_cons}. Now, for the virtual queue defined in \eqref{generic_virtual_queue}, it is easy to prove that
\begin{equation}\label{vq_UB}
    \frac{Z_{t+1}^2-Z^2_t}{2}\leq\frac{(x_{\max}-x_{\textrm{th}})^2}{2}+ Z_t(x_t-x_{\textrm{th}}),
\end{equation}
where $x_{\max}$ is the (finite by hypothesis) maximum value that $x_t$ can take, given system design constraints. Then, one can write the following upper bound \cite[Eqn. (4.47)]{Neely10}:
\begin{equation}\label{CDL_UB}
    \Delta_t(Z_t)\leq \frac{(x_{\max}-x_{\textrm{th}})^2}{2} +\mathbb{E}\left\{ Z_t(x_t-x_{\textrm{th}})|Z_t\right\}.
\end{equation}
Following stochastic optimization arguments \cite{Neely10}, it is sufficient to remove the expectation and minimize the CLD upper bound in \eqref{CDL_UB} in a per-slot basis, to guarantee the mean rate stability of $Z_t$ and, as a consequence, constraint \eqref{long_term_cons}, under the assumption of i.i.d. realizations of context parameters (an assumption that can be relaxed in some cases). However, in this way, no importance is assigned to the objective function of the problem, i.e., the goal cost in this case. To take the latter into account, denoting by $C_t$ the goal cost at time $t$, one can write the so called \textit{drift-plus-penalty} (DPP) function \cite{Neely10}:
\begin{equation}\label{DPP}    \Delta_p(Z_t)=\mathbb{E}\left\{L(Z_{t+1})-L(Z_t)+\Omega\cdot C_t|Z_t\right\},
\end{equation}
where $\Omega$ denotes a scalar hyper-parameter (the only one) controlling the trade-off between (original) constraint guarantees and objective function value minimization. The DPP is an augmented version of the CLD, and it penalizes high values of the objective function, e.g., excessive usage of resources in a network. Interestingly, by bounding the DPP, one can derive a similar result on virtual queue stability and long-term constraint guarantees, i.e., if $\Delta_p(Z_t)\leq B, \forall t$, the same result applies, with the following difference: by increasing the trade-off hyper-parameter $\Omega$, the value of the objective function decreases at the cost of longer convergence time of the virtual queue, and its average value \cite[Theorem 4.8]{Neely10}. Finally, to adapt the above analysis to problem \eqref{problem_formulation_EI}, we can define the following virtual queue evolution for constraint $(a)$ of problem \eqref{problem_formulation_EI}:
\begin{equation}\label{virtual_queue_EI}
    Z_{t+1}=\max\left(0,Z_t-\mathbf{1}_{\{\Theta_t\geq \Theta_{\textrm{th}}\}}\cdot\mathbf{1}_{\{D_{\textrm{tot},t}\leq D_{\max}\}}+\mathcal{E}_{g,th}\right).
\end{equation}
For virtual queue $Z_t$ in \eqref{virtual_queue_EI}, we can write (cf. \eqref{vq_UB}):
\begin{align}
    \frac{Z_{t+1}^2-Z_t^2}{2}\leq &\frac{(1-\mathcal{E}_{g,th})^2}{2}\nonumber\\
    &- Z_t\left(\mathbf{1}_{\{\Theta_t\geq \Theta_{\textrm{th}}\}}\cdot\mathbf{1}_{\{D_{\textrm{tot},t}\leq D_{\max}\}}-\mathcal{E}_{g,th}\right),\nonumber
\end{align}
which leads to the following DPP upper bound (recalling the objective function of \eqref{problem_formulation_EI}):
\begin{align}\label{DPP_UB_EI}
    \Delta_p(Z_t)\leq B_1 &+\mathbb{E}\big\{- Z_t\left(\mathbf{1}_{\{\Theta_t\geq \Theta_{\textrm{th}}\}}\cdot\mathbf{1}_{\{D_{\textrm{tot},t}\leq D_{\max}\}}\!\!-\!\mathcal{E}_{g,th}\right)\nonumber\\
    &-\Omega\cdot R_{d,t}|Z_t\big\},
\end{align}
with $B_1=\displaystyle\frac{(1-\mathcal{E}_{g,th})^2}{2}$. Finally, by minimizing \eqref{DPP_UB_EI} in a per-slot basis, we obtain the following instantaneous problem (we omit the constant terms):
\begin{align}\label{slot_problem}
    &\hspace{-1.5 cm}\underset{\mathbf{\Psi}_t}{\min}\quad - Z_t\mathbf{1}_{\{\Theta_t\geq \Theta_{\textrm{th}}\}}\cdot\mathbf{1}_{\{D_{\textrm{tot},t}\leq D_{\max}\}}-\Omega\cdot R_{d,t}\\
    &\textrm{subject to}\nonumber\\
    &(a)\; P_{d,t}\in\mathcal{P}_c,\;\quad (b)\; \gamma_t\in \Gamma_g.\nonumber
\end{align}
Hinging on the theoretical findings in \cite[Th. 4.5 and Th. 4.8]{Neely10}, under the i.i.d. assumption of context parameters (i.e., wireless channels, and remote computing delay), by solving \eqref{slot_problem} in each time slot, the mean-rate stability of the virtual queue is guaranteed (i.e., constraint $(a)$ of \ref{problem_formulation_EI} is met). Also, asymptotic optimality is achieved as the trade-off parameter $\Omega$ increases, at the cost of higher average virtual queue value and convergence time. Finally, thanks to the concept of $\Gamma$-additive approximation, non-exact solutions are allowed in expected sense, provided that the solution is within a bounded value $\Gamma$ from the infimum of all possible solutions, with an impact on the optimality performance of the algorithm that depends on $\Gamma$. Then, the next step is to solve problem \eqref{slot_problem}. Problem \eqref{slot_problem} is an integer program, however on a limited feasible set with cardinality $|\mathcal{P}_c|\times|\Gamma_g|$. As such, we will simply solve it through an exhaustive search over the feasible set. This is a typical outcome of Lyapunov stochastic optimization, thanks to the decoupling of the long-term problem into a sequence of (simpler) problems, based only on instantaneous observations of context parameters and properly defined state variables (virtual queues). The virtual queue (which evolves over time) and the parameter $\Omega$ drive the trade-off between $UE_d$ data rate loss (i.e., the goal cost), $UE_g$ goal-effectiveness, and convergence time.
\subsection{Solution of the instantaneous problem}\label{sec:solution}
\begin{algorithm}[t]
\SetAlgoLined
At each time slot $t$:\\
\begin{enumerate}
    \item Observe wireless channel realizations, computation resources, and virtual queue states;
    \item Solve \eqref{slot_problem} as described in Section \ref{sec:solution}
    \item Observe the real goal value outcome at the output of the classifier (\eqref{goal_value_EI}), and \\update the virtual queue $Z_t$ as in \eqref{virtual_queue_EI}
    \item Go to next time slot $t+1$
\end{enumerate}

\caption{Goal-oriented resource allocation}
\label{alg:GO_resource_allocation}
\end{algorithm}
Let us notice that the first indicator function in \eqref{slot_problem} ($\mathbf{1}_{\{\Theta_t\geq\Theta_{\textrm{th}}\}}$) is strongly dependent on the choice of the PER (equivalently, in practice, on the specific MCS employed for GO communication at time slot $t$). It should be noted that, in general, there is no known function relating PER and entropy relative increase (goal value). Also, differently from the one related to the delay (which is known if system state is perfectly known at the MEH - as we assume here), the actual value of such indicator variable is made available by the server only \textit{after the inference takes place}, i.e., after the computation phase, which is obviously too late to take a decision. Therefore, the decision has to be taken on a "best guess" of what could the outcome be. For this reason, as an approximation to solve \eqref{slot_problem}, we replace the first indicator function with its expectation, i.e. the probability of the goal value exceeding the threshold. Nevertheless, this approximation is not enough to solve the problem, as a model linking the PER and this probability is generally not available in advance. One possible solution is to exploit model-free optimization tools, such as DRL \cite{sutton2018reinforcement}. However, as the scope of this paper (and in particular of this section) is to provide a contribution to the coexistence of GO and DO systems, we will rely on a look up table built on a validation set. Specifically, using a validation set from the considered input pattern set, we build a look up table linking PER at $AP_g$ and the probability of the goal-value being above the predefined threshold, to be used during operation time to select the desired PER, based on current virtual queue states and context parameter realizations. 

Once the first indicator function in \eqref{slot_problem} (i.e., $\mathbf{1}_{\{\Theta_t\geq \Theta_{\textrm{th}}\}}$) is replaced with its expected value (i.e., the probability of the event) and the look up table is built, the remaining part of the problem can be \textit{optimally solved} via an exhaustive search over the involved variables. 
The overall procedure to dynamically select DO user transmit power and GO system PER is described in Algorithm \ref{alg:GO_resource_allocation}.


\section{Numerical evaluation}\label{sec:numer_eval}
In this section, we evaluate the performance of the proposed GO optimization strategy in Algorithm \ref{alg:GO_resource_allocation}, in the proposed coexistence network setting.
\begin{figure}[htb!]
    \centering
    \subfloat[Total system bandwidth $W=1$ GHz]{
        \includegraphics[width=\columnwidth]{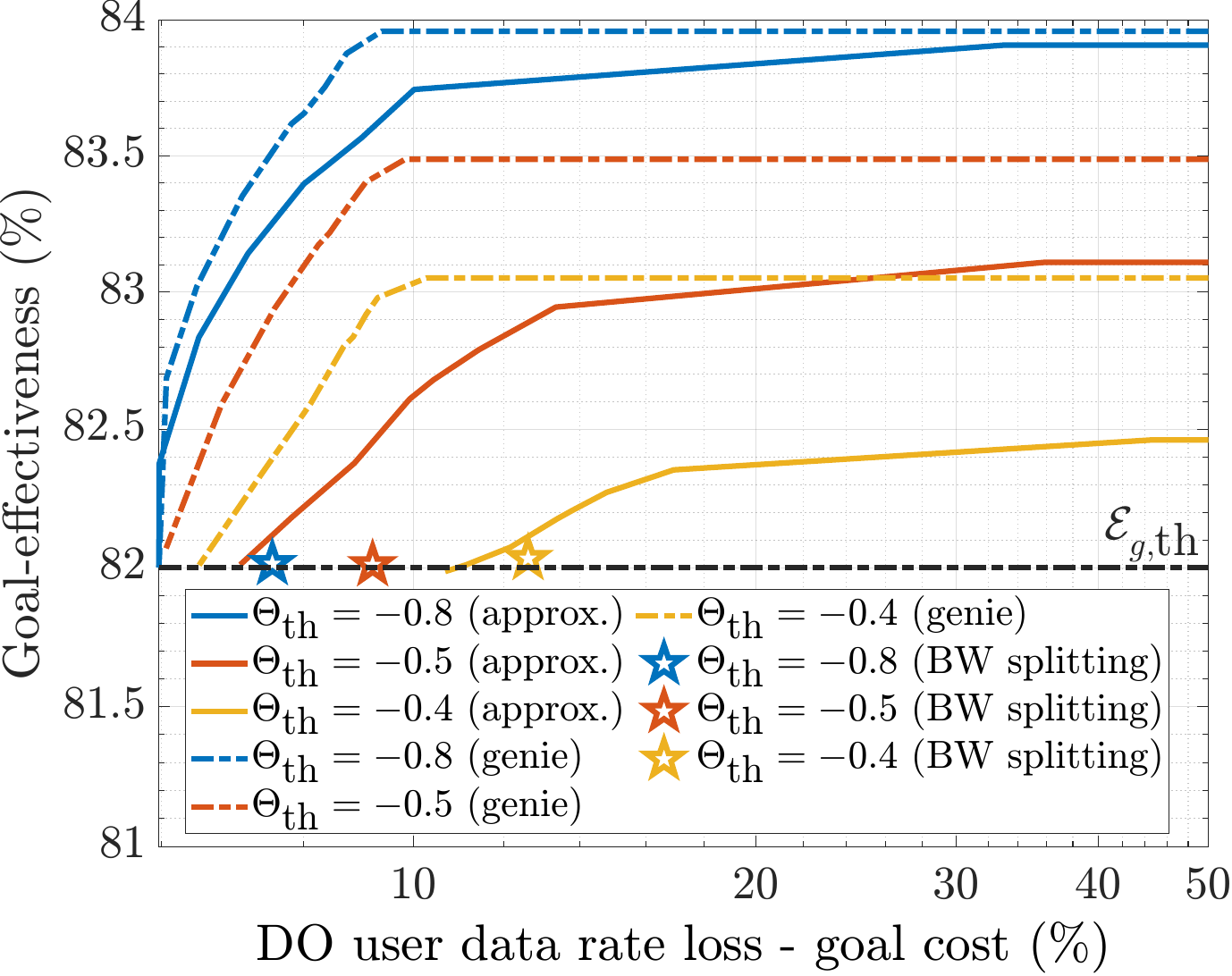}
        \label{fig:tradeoff1}
    }
    
    \subfloat[Total system bandwidth $W=500$ MHz]{
        \includegraphics[width=\columnwidth]{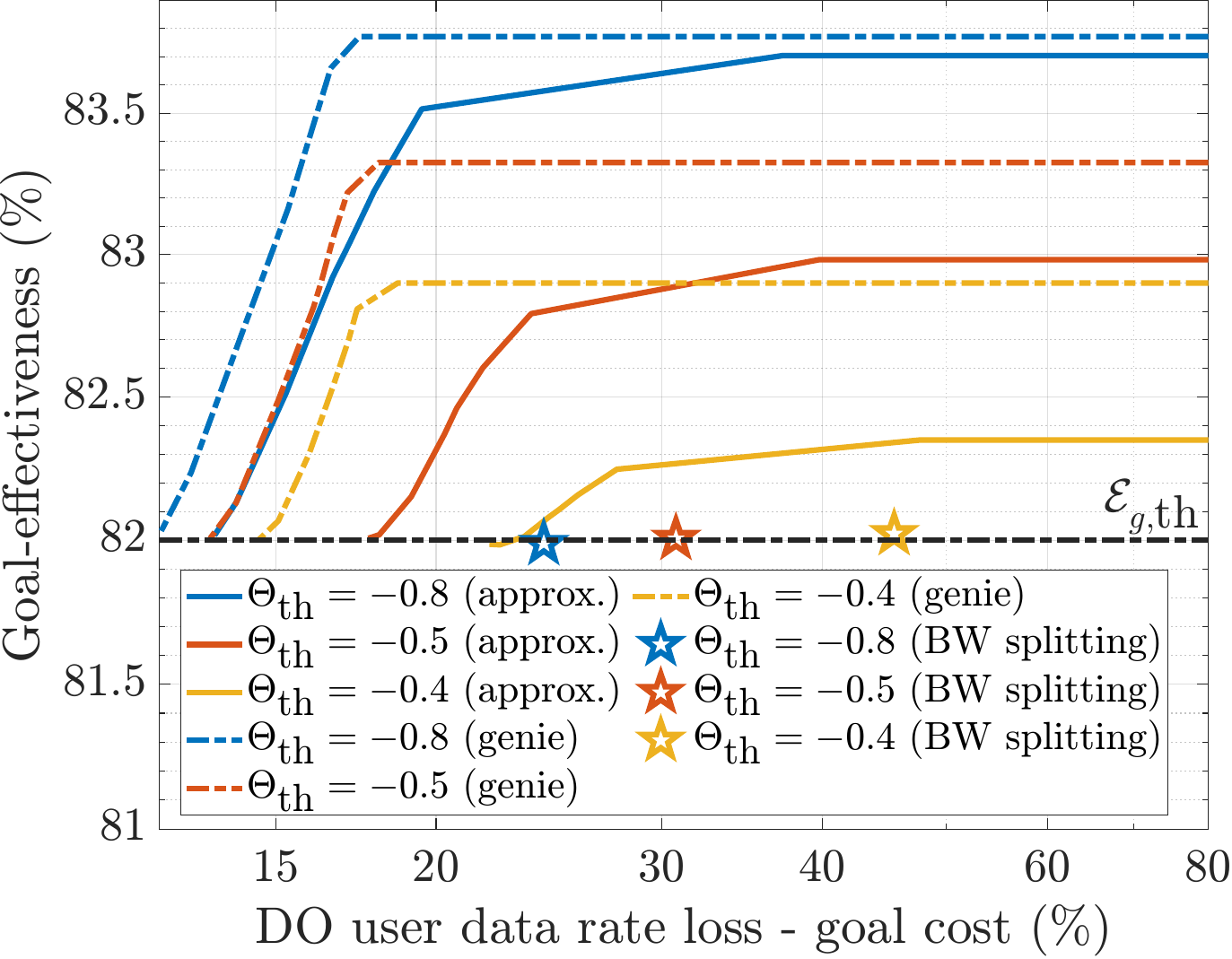}
        \label{fig:tradeoff1-b}
    }    
    \caption{Trade-off between goal cost and goal-effectiveness, with our method and with genie, against bandwidth splitting scenario}
    \label{fig:tradeoff}
\end{figure}
\begin{figure*}[htb!]
    \centering
    \subfloat[$\Theta_{\textrm{th}}=-0.8$]{
        \includegraphics[width=0.343\textwidth]{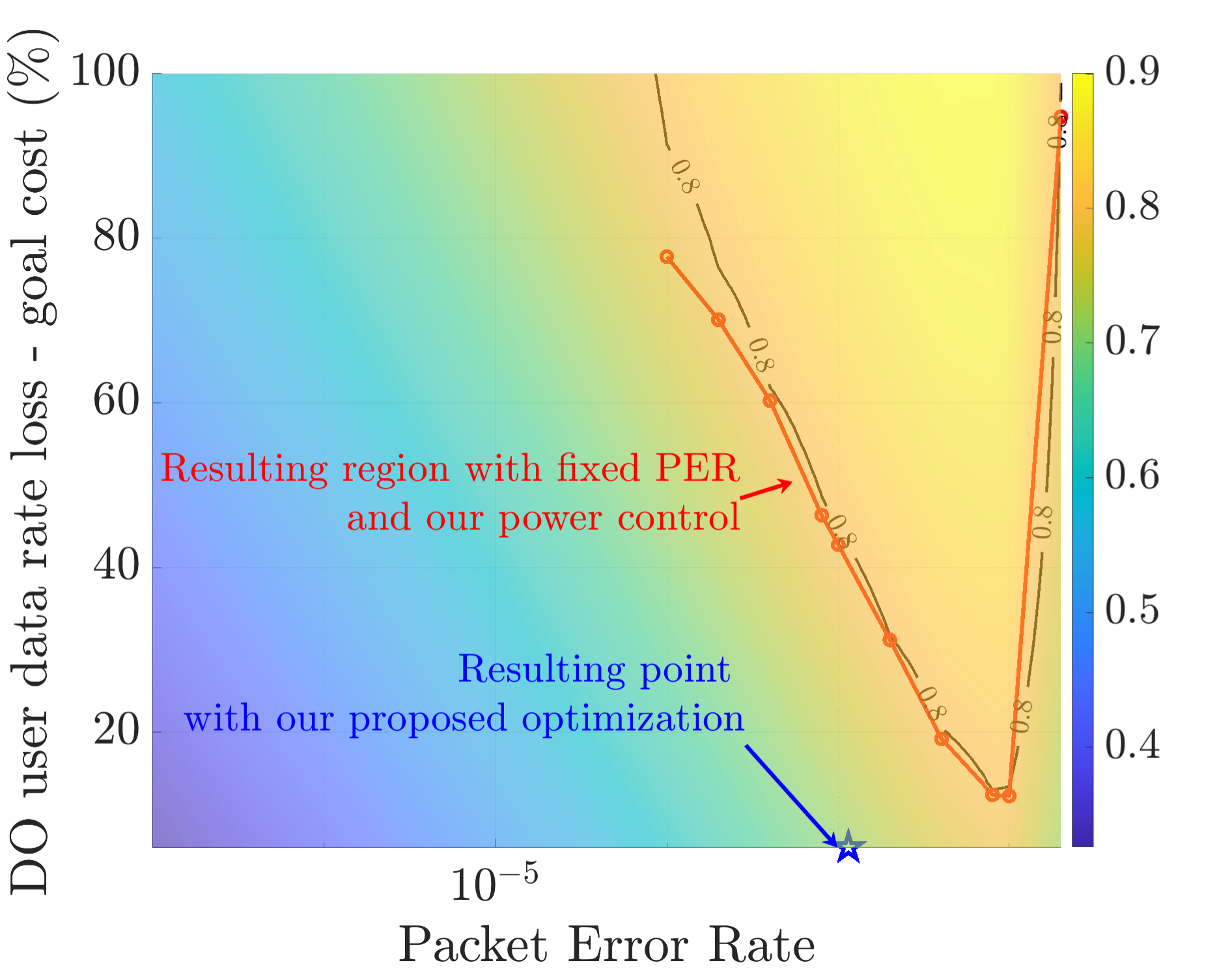}
        \label{fig:heatmap_entropy_80_vs_optimized}
    }
    \subfloat[$\Theta_{\textrm{th}}=-0.5$]{
        \includegraphics[width=0.343\textwidth]{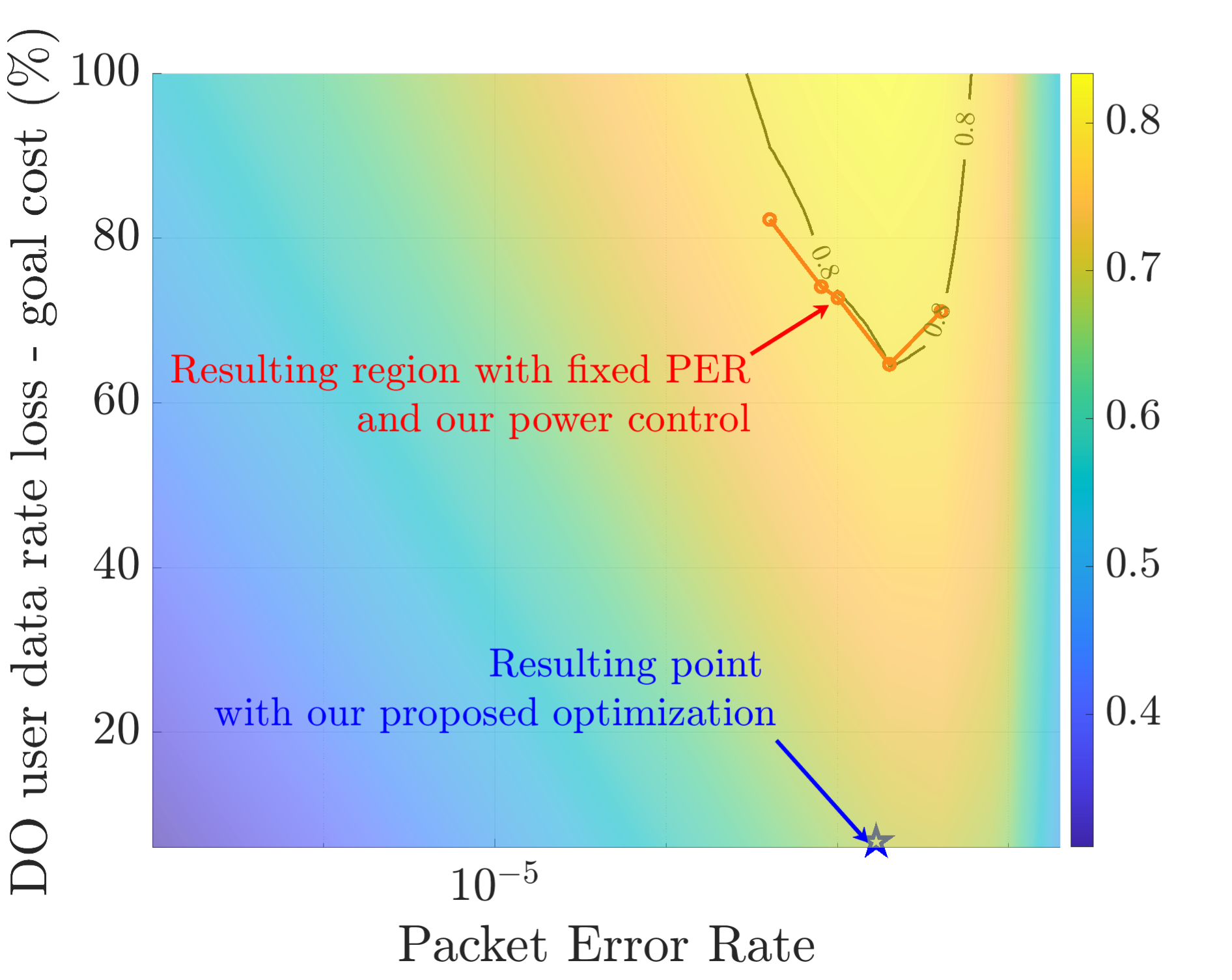}
        \label{fig:heatmap_entropy_50_vs_optimized}
    }
    \subfloat[$\Theta_{\textrm{th}}=-0.4$]{
        \includegraphics[width=0.343\textwidth]{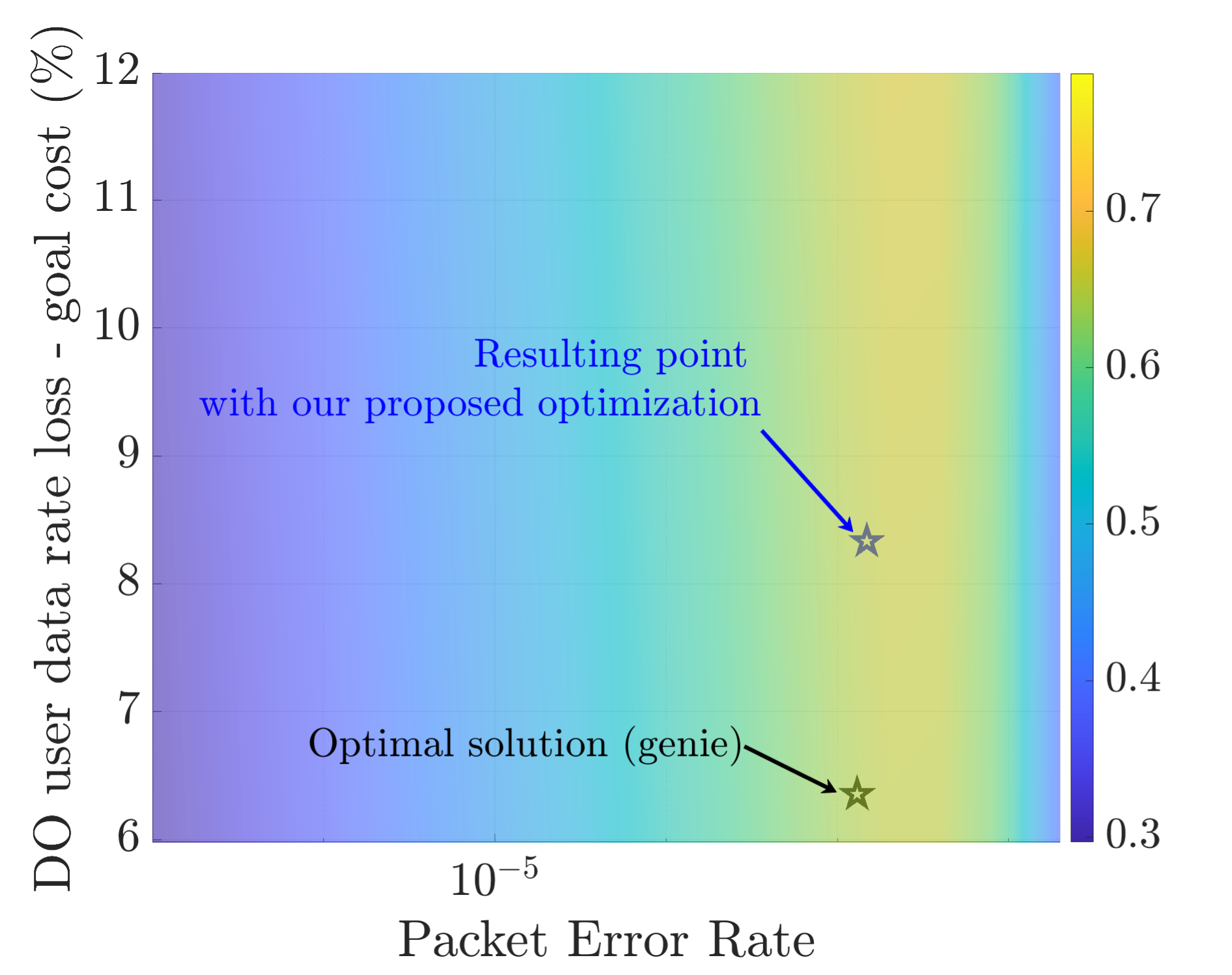}
        \label{fig:heatmap_entropy_40_vs_optimized}
    }
    \caption{Goal-effectiveness heat maps as a function of goal cost (DO user data rate loss) and target PER of GO system, with and without optimization, for fixed E2E delay threshold ($D_{\max}=45$ ms) and different goal value thresholds $\Theta_{\textrm{th}}$}
    \label{fig:heatmaps_vs_optimized}
    \vspace{-.5 cm}
\end{figure*}
\begin{figure}[htb!]
    \centering
    \subfloat[Temporal evolution of goal-effectiveness]{
        \includegraphics[width=.99\columnwidth]{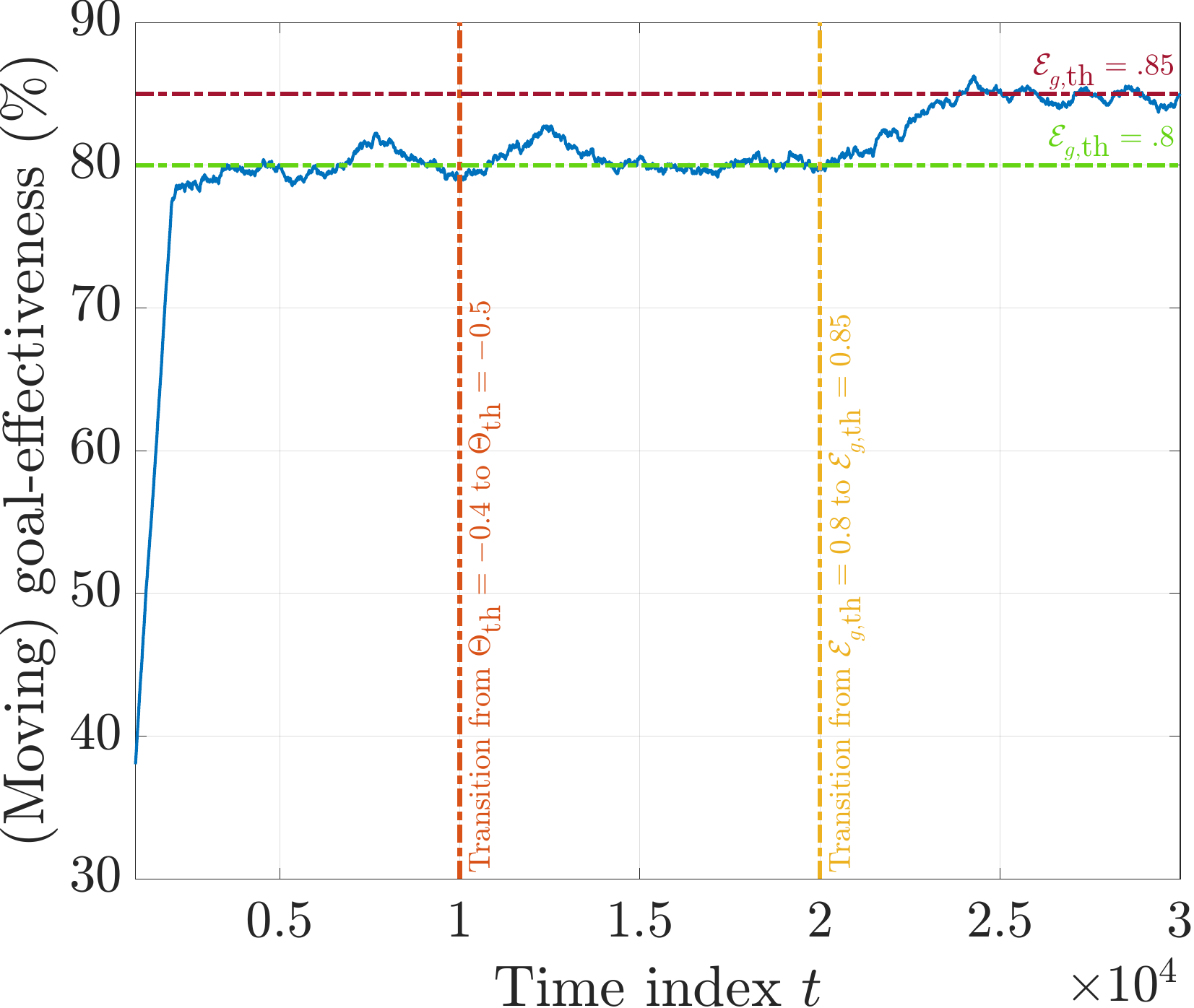}
        \label{fig:adaptation_goal_effectiveness}
    }
    
    \subfloat[Temporal evolution of goal cost]{
        \includegraphics[width=.99\columnwidth]{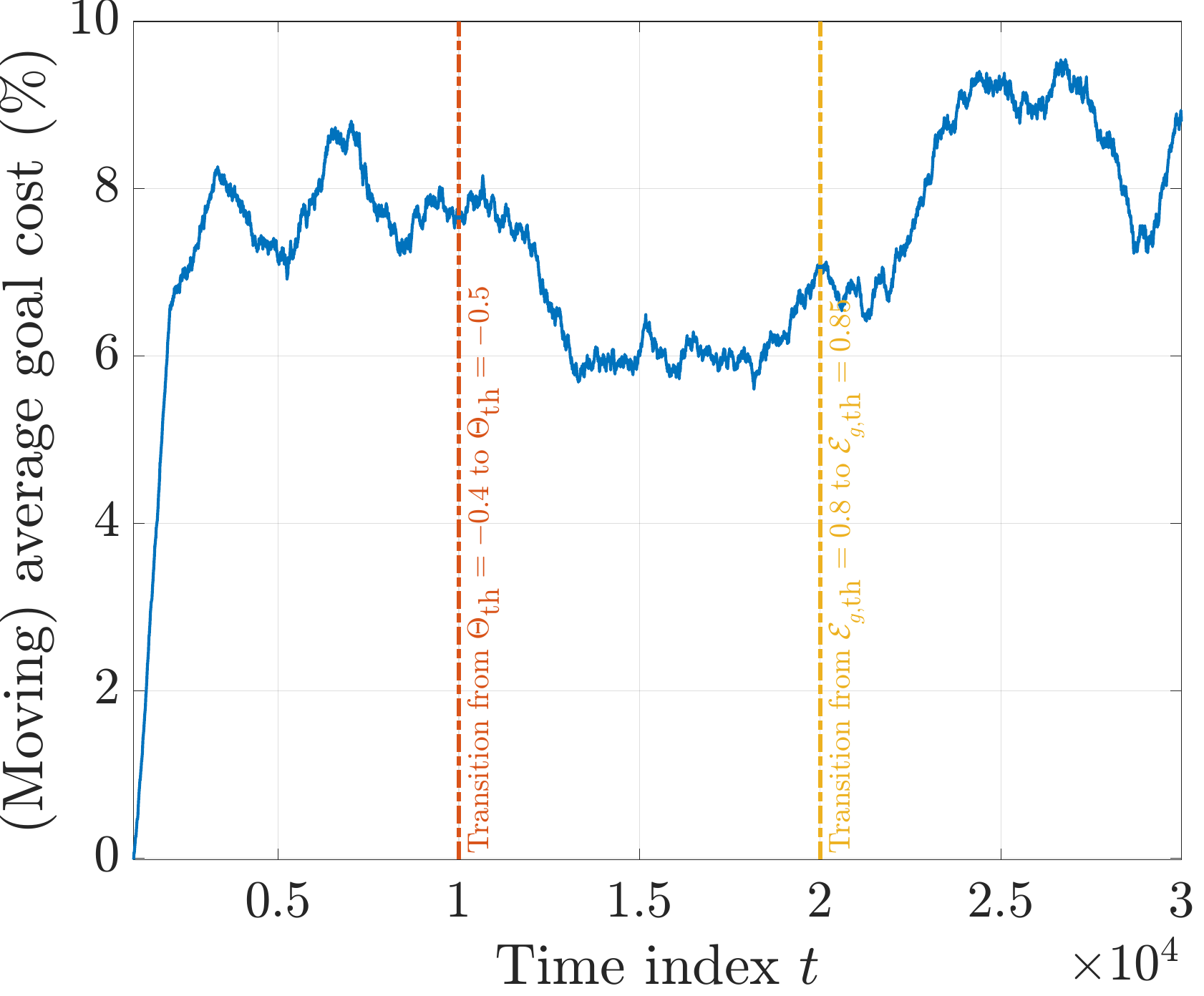}
        \label{fig:adaptation_goal_cost}
    }
    \caption{Adaptation capabilities of proposed method to online GO system requirement changes (i.e., $\Theta_{\textrm{th}}$ and $\mathcal{E}_g$)}
    \label{fig:adaptation1}
    \vspace{-.5 cm}
\end{figure}
\begin{figure}[htb!]
    \centering
    \subfloat[Temporal evolution of goal-effectiveness]{
        \includegraphics[width=\columnwidth]{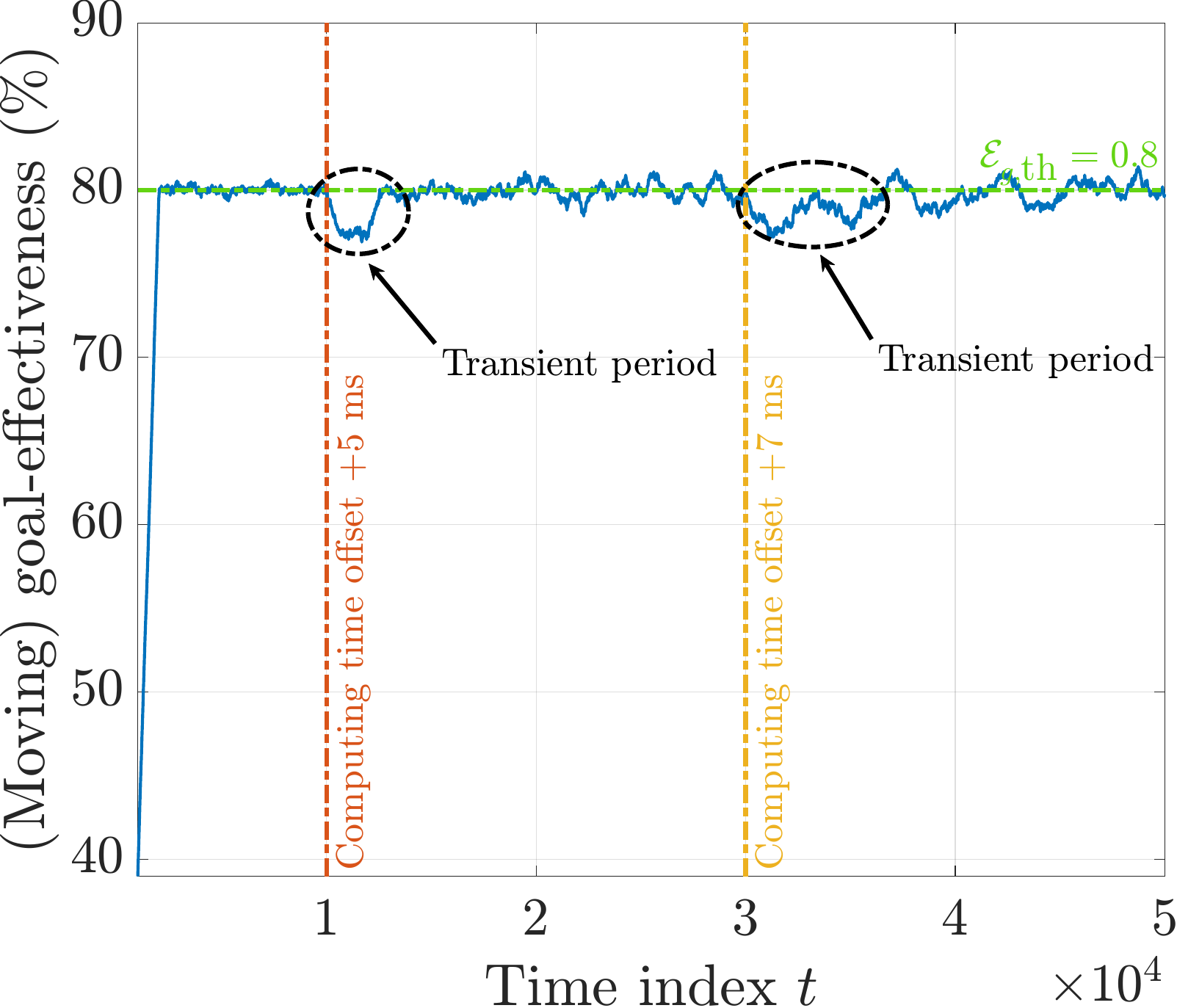}
        \label{fig:adaptation_goal_effectiveness_comp_time}
    }
    
    \subfloat[Temporal evolution of goal cost]{
        \includegraphics[width=\columnwidth]{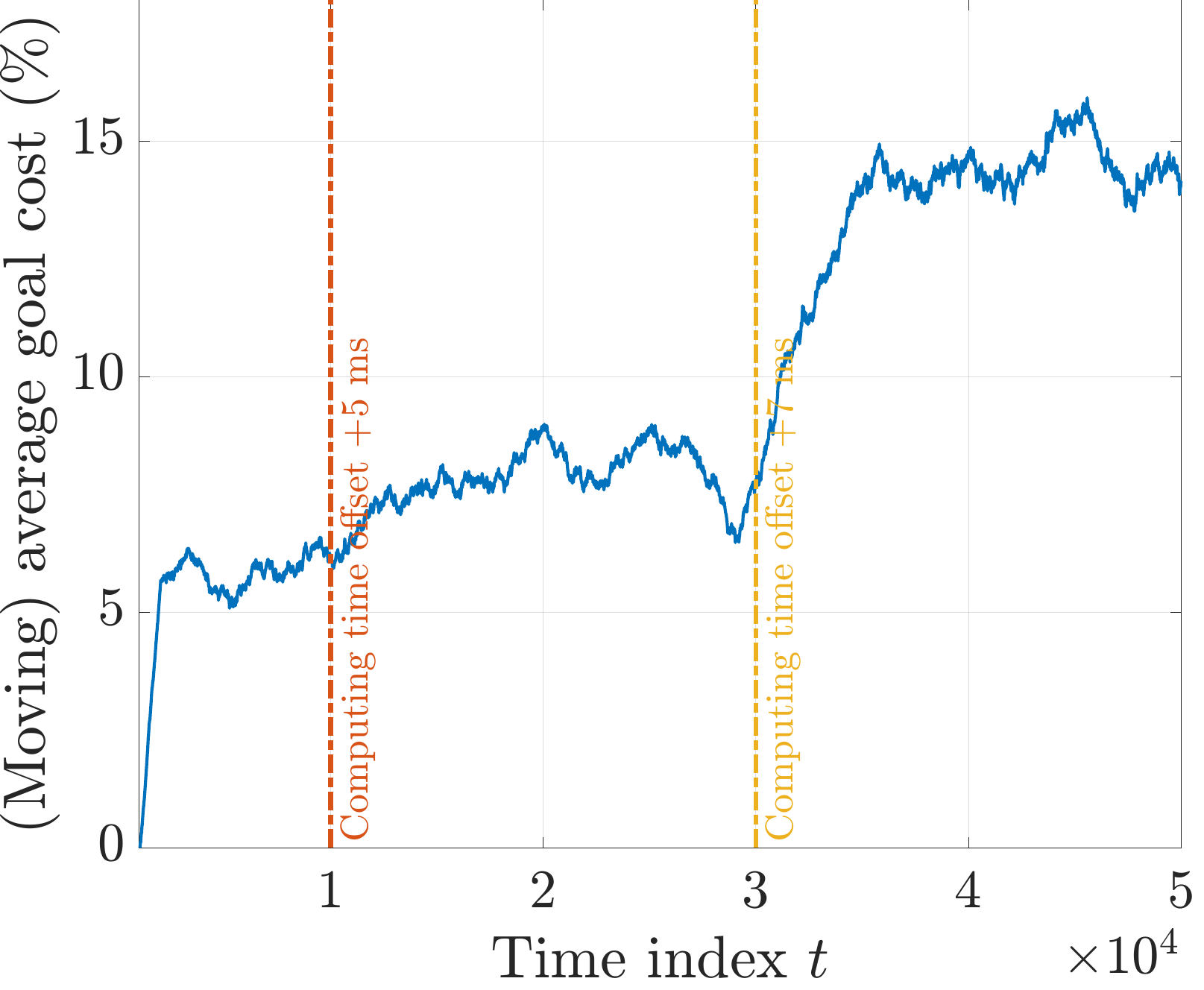}
        \label{fig:adaptation_goal_cost_comp_time}
    }
    \caption{Adaptation capabilities of proposed method to compute resource availability (i.e., computing delay)}
    \label{fig:adaptation2}
    \vspace{-.5 cm}
\end{figure}
To be able to compare the performance with fixed PER and $P_d$, we consider the same scenario of the previous evaluation (Fig. \ref{fig:heatmaps}). As a first result, in Fig. \ref{fig:tradeoff}, we show the trade-off between goal-effectiveness and goal cost, for a fixed goal-effectiveness threshold $\mathcal{E}_g=82$\%, two different values of total system bandwidth (Fig. \ref{fig:tradeoff1} with $W=1$ GHz and Fig. \ref{fig:tradeoff1-b} with $W=500$ MHz), and for different goal value thresholds (the same as the previous figure). Also, let us recall that our solution hinges on the problem approximation that replaces the first indicator function in \eqref{slot_problem} with its expectation computed on a validation set. Therefore, in Fig. \ref{fig:tradeoff1} and \ref{fig:tradeoff1-b}, we compare this solution with the one involving a \textit{genie} that uses a posteriori knowledge on the resulting entropy at the output of the classifier. Obviously, this solution cannot be implemented in practice, but it represent our benchmark, as it approaches the optimum of \eqref{problem_formulation_EI} as the trade-off parameter $\Omega$ increases. Finally, we compare our goal-oriented optimization method, with a strategy that splits the bandwidth across the two systems, thus resulting in zero co-channel interference leaked from one system to another, however, at the cost of less radio bandwidth available for each of the two systems. In this case, only the PER of the GO user is selected dynamically. This can also be considered as a goal-oriented communication KPIs selection, but the interference to and from a DO system is not managed, as the system is assumed to not be affected by interference. Note that the bandwidth splitting is empirically selected to obtain the target level of goal-effectiveness, to compare the results in terms of goal cost. In other words, $UE_g$ is allocated the minimum amount of bandwidth needed to achieve the target goal-effectiveness, while the remaining portion of the bandwidth is allocated to $UE_d$. In the curves of Fig. \ref{fig:tradeoff1}, the hyper-parameter $\Omega$ increases from right to left. The curves with the same colors are obtained with the same system parameters, but they represent the three different strategies: \textit{i)} GO optimization (i.e., our proposed strategy)  with solid lines, \textit{ii)} genie-aided GO optimization with dashed lines, and iii) bandwidth splitting scenario with the pentagrams. First, we can notice how our solution is able to always guarantee the goal-effectiveness constraint, with the goal cost decreasing towards the point in which the constraint is exactly attained (left hand side of the plot). Also, as the goal value threshold ($\Theta_{\textrm{th}}$) increases, the minimum achieved goal cost increases, as expected (see, e.g., yellow curve vs. blue curve). In particular, for the case with larger bandwidth (Fig. \ref{fig:tradeoff1}) the effect is more visible, but still limited, through the approximation, while the genie-aided solution is able to keep the cost close to optimal even when increasing $\Theta_{\textrm{th}}$. 
However, as $\Theta_{\textrm{th}}$ decreases, the approximated solution gets closer to the genie solution. Notably, \textit{both non-orthogonal spectrum sharing} solutions outperform the one with bandwidth splitting, with  considerable gain for the case of limited total available bandwidth (up to $50\%$ - Fig. \ref{fig:tradeoff1-b}). This suggests that the power of a GO system optimization is even more relevant under communication resource scarcity, as also pointed our in the introduction of this work. Still, it should be noted the large difference between the GO communication KPIs selection, and the legacy one, which is the purpose of the following performance evaluation.

As a second result, Fig. \ref{fig:heatmaps_vs_optimized} builds on the results of Fig. \ref{fig:heatmaps}, to compare our optimization method with the fixed PER and $P_d$ setting (i.e., the non optimized case). In particular, let us focus on Fig. \ref{fig:heatmap_entropy_80_vs_optimized}, i.e., the case with $\Theta_{\textrm{th}}=-0.8$. We compare three strategies: \textit{i)} the fixed (i.e., non adaptive) strategy of Fig. \ref{fig:heatmaps} (black curve), \textit{ii)} a strategy with fixed PER, with our adaptive algorithm only controlling $P_{d,t}$ (red curve), and \textit{iii)} our adaptive strategy for both PER and $P_d$ (blue pentagram). First, we can notice how strategy \textit{i)} (i.e., fixed PER and adaptive $P_{d,t}$) is shown only for a subset of possible PERs, i.e., only the feasible ones, given the target goal-effectiveness. Then, from the red curve, we can notice how the proposed optimization method (although without adaptive PER control) is able to follow the profile of the non optimized case, however slightly enlarging the feasibility region for some values of PER (e.g., $\gamma=10^{-4}$), i.e., achieving lower goal costs, thanks to power control and adaptation. However, this gain is negligible in most of the cases. On the other hand, the full optimization strategy (i.e., the adaptive PER and $P_{d}$ method), which is the main novelty of this work, achieves the lowest goal cost (around $6$\% vs. $12$\% of the fixed PER strategy), with a different resulting average PER. Similar considerations can be made for Fig. \ref{fig:heatmap_entropy_50_vs_optimized} (with $\Theta_{\textrm{th}}=-0.5$), however with a much larger gain obtained by our strategy with respect to the two benchmarks. Also, in Fig. \ref{fig:heatmap_entropy_40_vs_optimized}, the two non adaptive solutions do not provide feasible points for $\Theta_{\textrm{th}}=-0.4$, while our strategy is able to guarantee the target goal-effectiveness with a goal cost around $8.5$\%, to be compared with the optimal value obtained by the genie ($6.5$\%).

From Fig. \ref{fig:heatmaps_vs_optimized}, we can conclude that it is highly beneficial to adapt the target PER (i.e., communication reliability of GO communication) to higher layer performance (i.e., application performance) rather than adapting communication parameters to keep, e.g., a fixed target PER. This can have a huge consequence in adaptive MCS mechanisms, whose aim would be to guarantee application performance (adapting PER and interference tolerance) rather than guaranteeing a target PER \textit{a priori}, in the case of GO communications. 

Let us now focus on the adaptation capabilities of our proposed optimization. In particular, in Fig. \ref{fig:adaptation1}, we show a moving version of the goal-effectiveness (i.e., the goal-effectiveness estimated over the past $2000$ slots - Fig. \ref{fig:adaptation_goal_effectiveness}) and moving average goal cost (Fig. \ref{fig:adaptation_goal_cost}) as functions of time, introducing two unexpected events that need the algorithm to adapt to new conditions. Namely, at the beginning, the system requirements are: $D_{\max}=45$ ms, $\Theta_{\textrm{th}}=-0.4$ and $\mathcal{E}_g=0.8$. Then, after $10000$ iterations, we switch from $\Theta_{\textrm{th}}=-0.4$ to $\Theta_{\textrm{th}}=-0.5$, keeping the same target E2E delay and goal-effectiveness. After other $10000$ iterations (i.e., at slot $20000$), we switch from $\mathcal{E}_g=0.8$ to $\mathcal{E}_g=0.85$, while keeping the same goal value target and E2E delay requirement (i.e., $\Theta_{\textrm{th}}=-0.5$ and $D_{\max}=45$ ms). First, from Fig. \ref{fig:adaptation_goal_effectiveness}, we can notice how, from a goal-effectiveness perspective, the method is able to keep the desired value, switching to the new requirement after $t=20000$. Also, looking at Fig. \ref{fig:adaptation_goal_cost}, we can appreciate how the method always seeks for the optimal solution in terms of goal cost. Indeed, when switching from $\Theta_{\textrm{th}}=-0.4$ to $\Theta_{\textrm{th}}=-0.5$, the goal cost reduces, thanks to the more relaxed goal value threshold. Then, when imposing a stricter goal-effectiveness constraint ($t=20000$), the goal cost increases again to attain the desired performance. 

From Fig. \ref{fig:adaptation1}, it is then possible to appreciate the adaptation capabilities of the proposed method with respect to online requirement modifications. In particular, the method is able to autonomously detect this change through the virtual queue, and take corresponding actions to stabilize the system towards desired performance.

We have shown how the method behaves in the presence of requirement changes, but how does it react to a non stationary environment? Also, what is the effect of computation resource availability (i.e., computing delay) in the proposed scenario under investigation? 

To answer these questions, we consider a system in which, after $10000$ iterations, due to an exogenous event, the computing time experiences an offset of $+5$ ms, i.e., it is systematically increased by $5$ ms. Similarly, after $30000$ iterations, it is systematically increased by additional $2$ ms (i.e., $7$ ms with respect to the starting point). The E2E delay constraint is set to $D_{\max}=50$ ms. The results in terms of goal-effectiveness and goal cost are shown in Figs. \ref{fig:adaptation_goal_effectiveness_comp_time} and \ref{fig:adaptation_goal_cost_comp_time}, respectively, with the same approach of Fig. \ref{fig:adaptation1} (moving averages). From a goal-effectiveness perspective, we can notice how the method is able to attain the desired performance ($\mathcal{E}_g=80$\%), despite a transient period after the non stationarity appears in the system, highlighted by the dashed ellipses in the plot.  At the same time, adapting to the new system conditions (higher computing time) increases the cost, as a lower communication delay is required to guarantee the delay constraint (we always assume that the inference input/ image upload rate is such that no classification request buffering at the MEH occurs, even when the delay of GO communication is substantially reduced). Indeed, the only way is to reduce interference, i.e., the DO system data rate. Again, the method is able to autonomously detect this change through the virtual queue, and take corresponding actions to stabilize the system towards desired performance.

From Fig. \ref{fig:adaptation2}, we can conclude that \textit{the computing resources available to a GO system have a strong impact on the cost of a DO system that coexists and interferes with the former}. This is a clear way of seeing the effect of connect-compute resource availability in goal-oriented communication systems for edge inference.

From the numerical results shown in this section, we can draw the following general conclusions:
\begin{itemize}
    \item There exists a trade-off between goal-effectiveness and goal value, with the latter being related to communication performance of a DO system coexisting with the GO system; our method is able to explore this trade-off, with close to optimal performance in different conditions, depending on the specific requirements (cf. Fig. \ref{fig:tradeoff1}).
    \item Fixing the PER (i.e., adapting communication to maintain an a priori target PER), while adapting the DO user transmit power, does not provide much better performance than a strategy with both variables fixed across time (cf. \ref{fig:heatmaps_vs_optimized}). Higher gains are achieved via a fully adaptive system.
    \item Our method is able to dramatically reduce the goal cost, while guaranteeing target goal-effectiveness, by adaptively selecting target PER and DO user transmit power, based on measured application performance, even in the cases in which the fixed strategy fails to find a feasible solution (cf. Fig. \ref{fig:heatmap_entropy_40_vs_optimized}).
    \item Changing requirements over time (e.g., because of new application constraints) does not prevent our method from adaptively allocating resources to attain new levels of goal-effectiveness and/or goal values (cf. Fig. \ref{fig:adaptation1}). Also, the method works in both directions: it increases the cost when a transition to stricter requirements occurs, while it reduces the cost whenever requirements are relaxed. The latter, thanks to properly defined state variables (i.e., virtual queues) able to capture the behaviour of the system in terms of constraint violations. Obviously, this capability is limited to the cases in which non-stationarity occurs on a longer time scale than the method's adaptation.
    \item Computation resource availability at the GO system strongly affects the goal cost in terms of DO system data rate loss, a non trivial result, never shown in the literature before, to the best of our knowledge. Also non stationary environments, in terms of connect-compute resource availability, do not affect the adaptation capabilities of our method (cf. Fig. \ref{fig:adaptation2}).
\end{itemize}
\section{Conclusions}\label{sec:conclusions}

We have analyzed and optimized a wireless network scenario in which a GO and a legacy DO communication system coexist, fully sharing the same spectrum resources. While this has been proposed previously for the coexistence of data-oriented and semantic communication systems, we focused on GO communication, also analyzing the effect of communication errors and availability of computing resources. 

We first explained the concept of GO communications, and we provided a general problem formulation approach, to then tailor it to the proposed system scenario, showing how a GO resource allocation strategy can bring high gains in terms of overall system performance. The latter has been defined through two measures: goal-effectiveness and goal cost, with the former translating into probability of confident inference on time, and the latter referring to the performance loss of the legacy system, in terms of data rate. The proposed Go approach is to adapt communication KPIs (i.e., PER) to the actual outcome of communication, measured at the application level. Besides a first numerical performance evaluation for a non optimized setting, aimed at showing the potential trade-offs, our proposed optimization leverages on application performance measures to update suitably defined state variables, whose long-term stability has been exploited to achieve the goal with the lowest cost.  

After proposing an algorithm with theoretical guarantees, we tested it on the proposed scenario, in which the two systems coexist and interfere with each other. It has been shown, through numerical simulations, the gain, in terms of DO system data rate for a given goal-effectiveness of the GO system, of using an adaptive method for both communication reliability (i.e., PER), and legacy system user transmit power (i.e., affecting interference received by goal-oriented system and legacy system user data rate). Also, the proposed approach has been tested in non stationary environments, showing good adaptation capabilities to new requirements and computing resource availability, showing the strong link between communication and computing in future networks. Overall, the paper shows the superiority of a GO selection of communication KPIs, based on communication and computation resource availability, as well as interference.

Future steps involve scenarios with multiple GO and legacy system users, the optimization of GO user transmit power and other communication related parameters, cooperative inference, but also other applications beyond edge inference. Finally, an interesting research direction is that of semantic and GO feature extraction, into a unified framework in which only the most important features are transmitted and/or protected from wireless errors, by the GO users. From a methodological point of view, the exploration of data-driven techniques such as DRL could help solving the issue of not having a model relating communication reliability (e.g., PER) to the goal value. In this case, this relation would be learnt from experience, by the interaction of the agent(s) with the environment. Finally, we note that existing radio equipment, cybersecurity and privacy related regulation and market access frameworks such as the Radio Equipment Directive \cite{RegulationRED}, the EU Cybersecurity Act \cite{RegulationCyberAct} and the upcoming Cyber Resilience Act \cite{RegulationCRA} rely on classical communication 
approaches; new paradigms such as GO communications are considered insufficiently and corresponding adaptations will be necessary. 

\bibliographystyle{IEEEtran}
\bibliography{Main}

\begin{thebibliography}{10}
\providecommand{\url}[1]{#1}
\csname url@samestyle\endcsname
\providecommand{\newblock}{\relax}
\providecommand{\bibinfo}[2]{#2}
\providecommand{\BIBentrySTDinterwordspacing}{\spaceskip=0pt\relax}
\providecommand{\BIBentryALTinterwordstretchfactor}{4}
\providecommand{\BIBentryALTinterwordspacing}{\spaceskip=\fontdimen2\font plus
\BIBentryALTinterwordstretchfactor\fontdimen3\font minus
  \fontdimen4\font\relax}
\providecommand{\BIBforeignlanguage}[2]{{%
\expandafter\ifx\csname l@#1\endcsname\relax
\typeout{** WARNING: IEEEtran.bst: No hyphenation pattern has been}%
\typeout{** loaded for the language `#1'. Using the pattern for}%
\typeout{** the default language instead.}%
\else
\language=\csname l@#1\endcsname
\fi
#2}}
\providecommand{\BIBdecl}{\relax}
\BIBdecl

\bibitem{uusitalo2021}
M.~A. Uusitalo \emph{et~al.}, ``6{G} vision, value, use cases and technologies
  from european 6{G} flagship project {H}exa-{X},'' \emph{IEEE Access}, vol.~9,
  2021.

\bibitem{saad2020}
W.~Saad, M.~Bennis, and M.~Chen, ``A vision of 6{G} wireless systems:
  Applications, trends, technologies, and open research problems,'' \emph{IEEE
  Network}, vol.~34, no.~3, pp. 134--142, 2020.

\bibitem{hexa-XD1.2}
\BIBentryALTinterwordspacing
``Hexa-x deliverable {D}1.2 - {E}xpanded 6{G} vision, use cases and societal
  values – including aspects of sustainability, security and spectrum,''
  2021. [Online]. Available:
  \url{https://hexa-x.eu/wp-content/uploads/2021/05/Hexa-X_D1.2.pdf}
\BIBentrySTDinterwordspacing

\bibitem{hexa-XD1.3}
\BIBentryALTinterwordspacing
``{Hexa-X Deliverable {D}1.3 - Targets and requirements for 6{G} - initial E2E
  architecture},'' 2022. [Online]. Available:
  \url{https://hexa-x.eu/wp-content/uploads/2022/03/Hexa-X_D1.3.pdf}
\BIBentrySTDinterwordspacing

\bibitem{hexa-XD4.2}
\BIBentryALTinterwordspacing
``{Hexa-X Deliverable {D}4.2 - {A}{I}-driven communication \& computation
  co-design: initial solutions},'' 2022. [Online]. Available:
  \url{https://hexa-x.eu/wp-content/uploads/2022/07/Hexa-X_D4.2_v1.0.pdf}
\BIBentrySTDinterwordspacing

\bibitem{Strinati_RIS_2021}
E.~C. Strinati \emph{et~al.}, ``{Reconfigurable, Intelligent, and Sustainable
  Wireless Environments for 6G Smart Connectivity},'' \emph{IEEE Communications
  Magazine}, vol.~59, no.~10, pp. 99--105, 2021.

\bibitem{kekki2018etsi}
S.~Kekki, W.~Featherstone, Y.~Fang, P.~Kuure, A.~Li, A.~Ranjan, D.~Purkayastha,
  F.~Jiangping, D.~Frydman, G.~Verin \emph{et~al.}, ``{E}{T}{S}{I} {W}hite
  {P}aper: {M}{E}{C} in 5{G} networks,'' \emph{The European Telecommunications
  Standards Institute (ETSI), Tech. Rep. ETSI White Paper No. 28}, 2018.

\bibitem{Letaief2022}
K.~B. Letaief, Y.~Shi, J.~Lu, and J.~Lu, ``{Edge Artificial Intelligence for
  6G: Vision, Enabling Technologies, and Applications},'' \emph{IEEE Journal on
  Selected Areas in Communications}, vol.~40, no.~1, pp. 5--36, 2022.

\bibitem{Yang2020}
H.~Yang, A.~Alphones, Z.~Xiong, D.~Niyato, J.~Zhao, and K.~Wu,
  ``Artificial-intelligence-enabled intelligent 6{G} networks,'' \emph{IEEE
  Network}, vol.~34, no.~6, pp. 272--280, 2020.

\bibitem{Zhang2020}
\BIBentryALTinterwordspacing
S.~Zhang and D.~Zhu, ``Towards artificial intelligence enabled 6{G}: State of
  the art, challenges, and opportunities,'' \emph{Computer Networks}, vol. 183,
  p. 107556, 2020. [Online]. Available:
  \url{https://www.sciencedirect.com/science/article/pii/S138912862031207X}
\BIBentrySTDinterwordspacing

\bibitem{park2019wireless}
J.~Park, S.~Samarakoon, M.~Bennis, and M.~Debbah, ``Wireless network
  intelligence at the edge,'' \emph{Proceedings of the IEEE}, vol. 107, no.~11,
  pp. 2204--2239, 2019.

\bibitem{Zhou19_EI}
Z.~Zhou \emph{et~al.}, ``{Edge Intelligence: Paving the Last Mile of Artificial
  Intelligence With Edge Computing},'' \emph{Proceedings of the IEEE}, vol.
  107, no.~8, pp. 1738--1762, 2019.

\bibitem{Strinati_semantic2021}
E.~{Calvanese Strinati} and S.~Barbarossa, ``6{G} networks: Beyond {S}hannon
  towards semantic and goal-oriented communications,'' \emph{Computer
  Networks}, vol. 190, p. 107930, 2021.

\bibitem{Goldreich2012}
\BIBentryALTinterwordspacing
O.~Goldreich, B.~Juba, and M.~Sudan, ``A theory of goal-oriented
  communication,'' \emph{J. ACM}, vol.~59, no.~2, may 2012. [Online].
  Available: \url{https://doi.org/10.1145/2160158.2160161}
\BIBentrySTDinterwordspacing

\bibitem{ZouZhang2021}
H.~Zou, C.~Zhang, S.~Lasaulce, L.~Saludjian, and V.~Poor, ``Goal-oriented
  quantization: Analysis, design, and application to resource allocation,''
  \emph{https://arxiv.org/abs/2209.15347}, 09 2022.

\bibitem{binucci2022}
F.~Binucci, P.~Banelli, P.~Di~Lorenzo, and S.~Barbarossa, ``Dynamic resource
  allocation for multi-user goal-oriented communications at the wireless
  edge,'' in \emph{2022 30th European Signal Processing Conference (EUSIPCO)},
  2022, pp. 697--701.

\bibitem{ZhangZou2022}
C.~Zhang, H.~Zou, S.~Lasaulce, W.~Saad, M.~Kountouris, and M.~Bennis,
  ``Goal-oriented communications for the {I}o{T} and application to data
  compression,'' \emph{https://arxiv.org/abs/2211.05378}, 11 2022.

\bibitem{pezone2022}
F.~Pezone, S.~Barbarossa, and P.~Di~Lorenzo, ``Goal-oriented communication for
  edge learning based on the information bottleneck,'' in \emph{ICASSP 2022 -
  2022 IEEE International Conference on Acoustics, Speech and Signal Processing
  (ICASSP)}, 2022, pp. 8832--8836.

\bibitem{pappas2021}
N.~Pappas and M.~Kountouris, ``Goal-oriented communication for real-time
  tracking in autonomous systems,'' in \emph{2021 IEEE International Conference
  on Autonomous Systems (ICAS)}, 2021, pp. 1--5.

\bibitem{mostaani2019}
A.~Mostaani, O.~Simeone, S.~Chatzinotas, and B.~Ottersten, ``Learning-based
  physical layer communications for multiagent collaboration,'' in \emph{2019
  IEEE 30th Annual International Symposium on Personal, Indoor and Mobile Radio
  Communications (PIMRC)}, 2019, pp. 1--6.

\bibitem{Merluzzi_desiree}
M.~Merluzzi, A.~Martino, F.~Costanzo, P.~Di~Lorenzo, and S.~Barbarossa,
  ``Dynamic ensemble inference at the edge,'' in \emph{2021 IEEE Global
  Communications Conference (GLOBECOM)}, 2021, pp. 1--6.

\bibitem{Merluzzi2022_ICC}
M.~Merluzzi, C.~Battiloro, P.~Di~Lorenzo, and E.~C. Strinati,
  ``Energy-efficient classification at the wireless edge with reliability
  guarantees,'' in \emph{2022 IEEE International Conference on Communications
  Workshops (ICC Workshops)}, 2022, pp. 109--114.

\bibitem{MerluzziEML2021}
M.~Merluzzi, P.~Di~Lorenzo, and S.~Barbarossa, ``{Wireless Edge Machine
  Learning: Resource Allocation and Trade-Offs},'' \emph{IEEE Access}, vol.~9,
  pp. 45\,377--45\,398, 2021.

\bibitem{Shao2020}
J.~Shao and J.~Zhang, ``Bottlenet++: An end-to-end approach for feature
  compression in device-edge co-inference systems,'' in \emph{2020 IEEE
  International Conference on Communications Workshops (ICC Workshops)}, 2020,
  pp. 1--6.

\bibitem{Xie2022}
H.~Xie, Z.~Qin, X.~Tao, and K.~B. Letaief, ``Task-oriented multi-user semantic
  communications,'' \emph{IEEE Journal on Selected Areas in Communications},
  vol.~40, no.~9, pp. 2584--2597, 2022.

\bibitem{Sana2022}
M.~Sana and E.~C. Strinati, ``Learning semantics: An opportunity for effective
  6{G} communications,'' in \emph{2022 IEEE 19th Annual Consumer Communications
  \& Networking Conference (CCNC)}, 2022, pp. 631--636.

\bibitem{Lee19}
C.-H. Lee, J.-W. Lin, P.-H. Chen, and Y.-C. Chang, ``Deep learning-constructed
  joint transmission-recognition for internet of things,'' \emph{IEEE Access},
  vol.~7, pp. 76\,547--76\,561, 2019.

\bibitem{Merluzzi2022GO}
M.~Merluzzi, M.~C. Filippou, L.~G. Baltar, and E.~C. Strinati, ``Effective
  goal-oriented 6{G} communications: the energy-aware edge inferencing case,''
  in \emph{2022 Joint European Conference on Networks and Communications \& 6G
  Summit (EuCNC/6G Summit)}, 2022, pp. 457--462.

\bibitem{Mu2022}
X.~Mu and Y.~Liu, ``Exploiting semantic communication for non-orthogonal
  multiple access,'' \emph{https://arxiv.org/pdf/2209.06006.pdf}, 09 2022.

\bibitem{maman2021beyond}
M.~Maman \emph{et~al.}, ``{Beyond private 5{G} networks: applications,
  architectures, operator models and technological enablers},'' \emph{EURASIP
  Journal on Wireless Communications and Networking}, vol. 2021, no.~1, pp.
  1--46, 2021.

\bibitem{LEE2022}
\BIBentryALTinterwordspacing
H.~Lee, H.~Ahn, and Y.~D. Park, ``Performance analysis of coexistence of
  traditional communication system and emerging semantic communication
  system,'' \emph{ICT Express}, 2022. [Online]. Available:
  \url{https://www.sciencedirect.com/science/article/pii/S2405959522000571}
\BIBentrySTDinterwordspacing

\bibitem{kountouris2021}
M.~Kountouris and N.~Pappas, ``Semantics-empowered communication for networked
  intelligent systems,'' \emph{IEEE Communications Magazine}, vol.~59, no.~6,
  pp. 96--102, 2021.

\bibitem{Mu2022_conf}
X.~Mu and Y.~Liu, ``Semi-{N}{O}{M}{A} enabled coexisting semantic and bit
  communications,'' in \emph{2022 International Symposium on Wireless
  Communication Systems (ISWCS)}, 2022, pp. 1--6.

\bibitem{Hollnagel91}
E.~Hollnagel, ``A goals-means task analysis method,'' \emph{Available online:
  https://erikhollnagel.com/onewebmedia/GMTA.pdf}, 1991.

\bibitem{popovski14}
P.~Popovski, ``Ultra-reliable communication in 5{G} wireless systems,'' in
  \emph{1st International Conference on 5G for Ubiquitous Connectivity}, 2014,
  pp. 146--151.

\bibitem{MerluzziDMEC}
M.~Merluzzi, N.~d. Pietro, P.~Di~Lorenzo, E.~C. Strinati, and S.~Barbarossa,
  ``Discontinuous computation offloading for energy-efficient mobile edge
  computing,'' \emph{IEEE Transactions on Green Communications and Networking},
  vol.~6, no.~2, pp. 1242--1257, 2022.

\bibitem{BattiloroFL2022}
C.~Battiloro, P.~D. Lorenzo, M.~Merluzzi, and S.~Barbarossa, ``Lyapunov-based
  optimization of edge resources for energy-efficient adaptive federated
  learning,'' \emph{IEEE Transactions on Green Communications and Networking},
  pp. 1--1, 2022.

\bibitem{Sana19}
M.~Sana, A.~De~Domenico, W.~Yu, Y.~Lostanlen, and E.~Calvanese~Strinati,
  ``Multi-agent reinforcement learning for adaptive user association in dynamic
  mmwave networks,'' \emph{IEEE Transactions on Wireless Communications},
  vol.~19, no.~10, pp. 6520--6534, 2020.

\bibitem{Hao17}
H.~Yu and M.~J. Neely, ``Dynamic transmit covariance design in {M}{I}{M}{O}
  fading systems with unknown channel distributions and inaccurate channel
  state information,'' \emph{IEEE Transactions on Wireless Communications},
  vol.~16, no.~6, pp. 3996--4008, 2017.

\bibitem{Mao2016}
Y.~Mao, J.~Zhang, and K.~B. Letaief, ``Dynamic computation offloading for
  mobile-edge computing with energy harvesting devices,'' \emph{IEEE J. Sel.
  Areas Commun.}, vol.~34, no.~12, pp. 3590--3605, 12 2016.

\bibitem{Mao2017}
Y.~Mao, J.~Zhang, S.~H. Song, and K.~B. Letaief, ``Stochastic joint radio and
  computational resource management for multi-user mobile-edge computing
  systems,'' \emph{IEEE Trans. Wireless Commun.}, vol.~16, no.~9, pp.
  5994--6009, 9 2017.

\bibitem{Merluzzi2020URLLC}
M.~{Merluzzi}, P.~{Di Lorenzo}, S.~{Barbarossa}, and V.~{Frascolla}, ``{Dynamic
  Computation Offloading in Multi-Access Edge Computing via Ultra-Reliable and
  Low-Latency Communications},'' \emph{IEEE Transactions on Signal and
  Information Processing over Networks}, pp. 1--1, 2020.

\bibitem{Shlezinger2021}
N.~Shlezinger, E.~Farhan, H.~Morgenstern, and Y.~C. Eldar, ``Collaborative
  inference via ensembles on the edge,'' in \emph{ICASSP 2021 - 2021 IEEE
  International Conference on Acoustics, Speech and Signal Processing
  (ICASSP)}, 2021, pp. 8478--8482.

\bibitem{Hexa-X2021}
\BIBentryALTinterwordspacing
Hexa-X. (2021, Apr.) {D}1.2 -- expanded 6g vision, use cases and societal
  values –- including aspects of sustainability, security and spectrum.
  [Online]. Available:
  \url{https://hexa-x.eu/wp-content/uploads/2021/05/Hexa-X_D1.2.pdf}
\BIBentrySTDinterwordspacing

\bibitem{mitola1999cognitive}
J.~Mitola and G.~Q. Maguire, ``Cognitive radio: making software radios more
  personal,'' \emph{IEEE personal communications}, vol.~6, no.~4, pp. 13--18,
  1999.

\bibitem{haykin2005cognitive}
S.~Haykin, ``Cognitive radio: brain-empowered wireless communications,''
  \emph{IEEE journal on selected areas in communications}, vol.~23, no.~2, pp.
  201--220, 2005.

\bibitem{Polyanskiy2010}
Y.~Polyanskiy, H.~V. Poor, and S.~Verdu, ``Channel coding rate in the finite
  blocklength regime,'' \emph{IEEE Transactions on Information Theory},
  vol.~56, no.~5, pp. 2307--2359, 2010.

\bibitem{Sun19}
C.~Sun, C.~She, C.~Yang, T.~Q.~S. Quek, Y.~Li, and B.~Vucetic, ``Optimizing
  resource allocation in the short blocklength regime for ultra-reliable and
  low-latency communications,'' \emph{IEEE Transactions on Wireless
  Communications}, vol.~18, no.~1, pp. 402--415, 2019.

\bibitem{She17}
C.~She, C.~Yang, and T.~Q.~S. Quek, ``Radio resource management for
  ultra-reliable and low-latency communications,'' \emph{IEEE Communications
  Magazine}, vol.~55, no.~6, pp. 72--78, 6 2017.

\bibitem{Canziani16}
E.~C. A.~Canziani, A.~Paszke, ``An analysis of deep neural network models for
  practical applications,'' \emph{Available online:
  https://arxiv.org/abs/1605.07678}, 2016.

\bibitem{bishop2006pattern}
C.~M. Bishop, \emph{Pattern recognition and machine learning}.\hskip 1em plus
  0.5em minus 0.4em\relax Springer, 2006.

\bibitem{Saerens2002}
M.~Saerens, P.~Latinne, and C.~Decaestecker, ``Any reasonable cost function can
  be used for a posteriori probability approximation,'' \emph{IEEE Transactions
  on Neural Networks}, vol.~13, no.~5, pp. 1204--1210, 2002.

\bibitem{krizhevsky2009learning}
A.~Krizhevsky and G.~Hinton, ``Learning multiple layers of features from tiny
  images,'' University of Toronto, Toronto, Ontario, Tech. Rep.~0, 2009.

\bibitem{Neely10}
M.~J. Neely, \emph{Stochastic Network Optimization with Application to
  Communication and Queueing Systems}.\hskip 1em plus 0.5em minus 0.4em\relax
  Morgan \& Claypool Publishers, 2010.

\bibitem{sutton2018reinforcement}
R.~S. Sutton and A.~G. Barto, \emph{Reinforcement learning: An
  introduction}.\hskip 1em plus 0.5em minus 0.4em\relax MIT press, 2018.

\bibitem{RegulationRED}
``Directive 2014/53/{E}{U} of the {E}uropean {P}arliament and of the {C}ouncil
  of 16 april 2014 on the harmonisation of the laws of the {M}ember {S}tates
  relating to the making available on the market of radio equipment and
  repealing {D}irective 1999/5/{E}{C},'' 2014.

\bibitem{RegulationCyberAct}
``Regulation ({E}{U}) 2019/881 of the european parliament and of the council of
  17 april 2019 on {E}{N}{I}{S}{A} (the {E}uropean {U}nion {A}gency for
  {C}ybersecurity) and on information and communications technology
  cybersecurity certification and repealing {R}egulation ({E}{U}) no 526/2013
  ({C}ybersecurity {A}ct),'' 2019.

\bibitem{RegulationCRA}
``Proposal for a {R}egulation of the {E}uropean {P}arliament and of the
  {C}ouncil on horizontal cybersecurity requirements for products with digital
  elements and amending {R}egulation ({E}{U}) 2019/1020,'' 2022.

\end{thebibliography}

\vfill\pagebreak

\end{document}